\documentclass[12pt,notoc]{JHEP3}

\usepackage{amsmath,amssymb,euscript,array,cite,mathrsfs}

\def\newf{{\cal F}}
\def\oldf {f}

\newcommand{\startappendix}{
\setcounter{section}{0}
\renewcommand{\thesection}{\Alph{section}}}
\newcommand{\Appendix}[1]{
\refstepcounter{section}
\begin{flushleft}
{\large\bf Appendix \thesection: #1}
\end{flushleft}}
\usepackage{epsfig}

\def\C{\Gamma}

\newcommand{\sech}{\operatorname{sech}}

\newcommand{\Tr}{\operatorname{Tr}}
\def\BI{{\boldsymbol I}}
\def\B0{{\boldsymbol 0}}
\def\BF{{\boldsymbol F}}
\def\BK{{\boldsymbol K}}
\def\BX{{\boldsymbol X}}
\def\BZ{{\boldsymbol Z}}
\def\Bp{{\boldsymbol p}}
\def\Bw{{\boldsymbol w}}
\def\BOmega{{\boldsymbol\Omega}}
\def\Bvarpi{{\boldsymbol\varpi}}
\def\Bpi{{\boldsymbol\pi}}
\def\CP{{\mathbb C}P}
\def\RP{{\mathbb R}P}

\def\Tr{{\rm Tr}}

\def\det{{\rm det}}

\def\C{{\mathbb C}}

\def\ee{\boldsymbol{e}}

\def\R{{\mathbb R}}

\newcommand{\BH}{\boldsymbol{H}}

\def\Dbarslash{\,\,{\raise.15ex\hbox{/}\mkern-12mu {\bar D}}}
\def\Dslash{\,\,{\raise.15ex\hbox{/}\mkern-12mu D}}
\def\delslash{\,\,{\raise.15ex\hbox{/}\mkern-9mu \partial}}
\def\delbarslash{\,\,{\raise.15ex\hbox{/}\mkern-9mu {\bar\partial}}}

\def\ms{{\mathfrak M}}

\def\Q{{\cal Q}}

\def\LAG{\mathscr{L}}

\newcommand{\MAT}[1]{\begin{pmatrix} #1\end{pmatrix}}
\newcommand{\EQ}[1]{\begin{equation} #1 \end{equation}}
\newcommand{\AL}[1]{\begin{subequations}\begin{align} #1
\end{align}\end{subequations}}
\newcommand{\SP}[1]{\begin{equation}\begin{split} #1
\end{split}\end{equation}}

\title{Magnons, their Solitonic Avatars and the Pohlmeyer Reduction}
\author{Timothy J. Hollowood\\
Department of Physics,\\ University of Wales Swansea,\\
Swansea, SA2 8PP, UK.\\
E-mail: \email{t.hollowood@swansea.ac.uk}}

\author{and J.~Luis Miramontes\\
Departamento de F\'\i sica de Part\'\i culas and IGFAE,\\
Universidad
de Santiago de Compostela\\ 15782 Santiago de Compostela, Spain\\
E-mail: \email{jluis.miramontes@usc.es}}

\abstract{We study the solitons of the symmetric space sine-Gordon
  theories that arise once the Pohlmeyer reduction has been imposed
  on a sigma model with the symmetric space as target. Under this map
  the solitons arise as giant magnons that are relevant to string
  theory in the context of the AdS/CFT correspondence. In particular,
  we consider the cases $S^n$, $\CP^n$ and $SU(n)$ in some detail.
We clarify the construction of the charges
  carried by the solitons and also address the possible Lagrangian
  formulations of the symmetric space sine-Gordon theories. 
We show that the dressing, or B\"acklund,
  transformation naturally produces solitons directly in both the sigma
  model and the symmetric space sine-Gordon equations without the need
  to explicitly map from one to the other. In particular, we obtain a
  new magnon solution in $\CP^3$.
We show that the dressing method does not 
produce the more general ``dyonic'' solutions which involve
  non-trivial motion of the collective coordinates carried by the solitons. 
}

\begin{document}

\section{Introduction}

The AdS/CFT correspondence~\cite{Maldacena:1997re}
 is remarkable in so many ways. For example,
there is an underlying integrable structure that allows
one to interpolate from weak to strong coupling, and which enables many quantitave checks of the conjectured duality by exploring both sides of the correspondence~(see~\cite{Reviews} and the references therein).
An example is provided by the ``giant magnons'' and the ``dyonic giant
 magnons'', introduced by Hofman and Maldacena~\cite{Hofman:2006xt}
 and Dorey~\cite{DyonicGM}, respectively. They describe string
 configurations on curved space-times of the form ${\mathbb R}_t\times \ms$, with $\ms=F/G$ a symmetric space; for example, $S^n=SO(n+1)/SO(n)$.
Then, the classical motion of the string is described by a sigma model with target space $\ms$, and the Virasoro constraints, in a particular
gauge, lead to the Pohlmeyer reduction of that sigma model~\cite{Pohlmeyer:1975nb,Tseytlin:2003ii}. In turn, this gives rise to an associated integrable system that is a
generalization of the sine-Gordon theory. These are the symmetric
space sine-Gordon theories (SSSG), and giant magnons can be mapped into the soliton solutions to their equations-of-motion. 
Moreover, when the symmetric space is of indefinite signature, like $AdS_n=SO(2,n-1)/SO(1,n-1)$, similar ideas can be used to study also the motion of strings on curved spaces of the form $\ms\times S^1$, or even $\ms$.
In the context of the AdS/CFT correspondence, giant magnons have been
 extensively used to study many aspects of superstrings in certain
 subspaces of $AdS_5\times
 S^5$~\cite{Hofman:2006xt,Gubser:1998bc,Frolov:2002av} and 
$AdS_4\times \CP^3$~\cite{Gaiotto:2008cg,Grignani:2008is,Abbott:2008qd}.

This work is a companion to~\cite{Miramontes:2008wt}, which 
provided a systematic study of the group theoretical interpretation of
the Pohlmeyer reduction and the associated SSSG
theories for symmetric spaces of definite, or indefinite, signature. The
present work extends this to a discussion and construction of a class
of soliton solutions using the dressing 
transformation method~\cite{Dressing}.
An important result that we establish is that the
dressing method produces both the giant magnon {\it and\/} its soliton
avatar in the SSSG theory at the same time, without the need to map one
to other via the Pohlmeyer constraints. This is particularly useful
because, in general, it is not easy to perform the map.

For cases including $S^n$ the giant magnon solutions produced by the
dressing method have been studied in
\cite{Spradlin:2006wk} and they correspond to
embeddings of the 
Hofman--Maldacena giant magnon  \cite{Hofman:2006xt} associated to
$S^2\subset S^5$. One major shortcoming of the dressing
method is that, in the context of $S^n=SO(n+1)/SO(n)$, it does not
produce the Dorey's dyonic giant magnon
\cite{DyonicGM}. Nevertheless, this more general
solution can be constructed by the dressing procedure by using the
alternative formulation of $S^3$ as the symmetric space $SU(2)\times
SU(2)/SU(2)$, which is isomorphic to the Lie group $SU(2)$. Embedding this
solution back in the 
original formulation in terms of the symmetric space
$S^n=SO(n+1)/SO(n)$ 
shows that the dyonic solution involves non-trivial
geodesic motion in the space of collective coordinates carried by the 
magnon/soliton. In this sense, the solutions have much in common with
the dyonic generalization of the monopole in four-dimensional gauge
theories coupled to an adjoint Higgs field~(for example see
\cite{Gibbons:1986df}.

Other examples that we consider in this work are the complex
projective spaces $\CP^n$, which are realized as the symmetric spaces
$SU(n+1)/U(n)$. The case $\CP^3$ is relevant to the  
AdS/CFT correspondence involving a spacetime
$AdS_4\times\CP^3$~\cite{NewDuality}. The known giant magnon solutions for this case
\cite{Gaiotto:2008cg,Grignani:2008is,Abbott:2008qd,Astolfi:2008ji,Grignani:2008te}
have all been obtained from the Hofman-Maldacena solution
and the dyonic generalization of Dorey 
via embeddings of $S^2$ and $S^3$ in $\CP^n$, respectively.
Our results provide a class of new magnon/soliton solutions
which cannot be obtained from embeddings of those for $S^n$. In
addition, we show that there should exist an equivalent class of
dyonic solutions in addition to the embeddings of Dorey's dyon. 

Finally, we consider the $SU(n)$ principal chiral models, which can be
formulated as a symmetric space $SU(n)\times SU(n)/SU(n)$. 
For $n>2$, these models admit several non-equivalent Pohlmeyer reductions and, therefore, they give rise to different SSSG theories. In this work we only consider the simplest cases, which correspond to the  (parity invariant) homogeneous sine-Gordon theories~\cite{HSG}. The solitons of these theories have been studied in~\cite{FernandezPousa:1997iu} using a different formulation of the dressing transformation method based on representations of affine Lie algebras. Our results provide new expressions for them involving collective coordinates that clarify their composite nature in terms of basic $SU(2)$ solitons.
More general reductions of the principal chiral models will be discussed elsewhere.

Notice that all the examples that we consider 
involve symmetric spaces of definite signature. In future work we
will describe the generalization to symmetric spaces of indefinite
signature relevant to discussing $AdS_n$, for example.

The plan of the paper is a follows. In Section~\ref{SSSM}, we will formulate the sigma model with target space a symmetric space $F/G$
in terms of a constrained $F$-valued field without introducing gauge fields. The relationship between this formulation and the approach used in~\cite{Miramontes:2008wt} is summarized in Appendix~\ref{AppSigmaModel}. Using that formulation, in Section~\ref{Pohl}
we will describe the Pohlmeyer reduction of the sigma model, and recover the formulation of the SSSG equations as zero-curvature conditions on a left-right asymmetric coset of the form $G/H^{(-)}_L\times H^{(+)}_R$ proposed in~\cite{Miramontes:2008wt}. We will also address the possible Lagrangian formulations of these equations and clarify their symmetries and conserved quantities, which play an important r\^ole in the description of the soliton solutions. 
In Section~\ref{GM}, we will review the already known giant magnons in
the context of $S^n$ and $\CP^n$, and we will discuss the relation between
them and their relativistic SSSG solitonic avatars. In
Section~\ref{dress}, we will use the dressing transformation method to
construct magnons and solitons following the approach
of~\cite{Harnad:1983we}. An important result of this section is that
the dressing transformation is compatible with the Pohlmeyer
reduction, and that this method provides directly both the magnon and
its SSSG soliton without the need to map one into the other. In
Section~\ref{PCM}, we will apply the method to the $SU(n)$ principal
chiral model, and for $n=2$ we will recover Dorey's dyonic giant
magnon. In Sections~\ref{CPn} and~\ref{Sn}, we will apply the method
to $\CP^n$ and $S^n$, respectively. In Section~\ref{Beyond}, we
discuss the possibility of finding solutions similar to Dorey's dyonic
giant magnon by making the collective coordinates time
dependent. Finally, Section~\ref{Conclusions} contains our conclusions,
and there are four appendices. 

\section{Symmetric Space Sigma Model}
\label{SSSM}

Our story begins with a sigma model in $1+1$ dimensions whose target
space is a symmetric space, that is a quotient of two Lie groups 
$F/G$ equipped with an involution $\sigma_-$ of $F$ that fixes $G\subset F$:
\EQ{
\sigma_-(g)=g\ ,\qquad\forall g\in G\ .
}
Acting on ${\mathfrak f}$, the Lie algebra of $F$, the automorphism $\sigma_-$ gives rise to the canonical orthogonal decomposition 
\begin{equation}
{\mathfrak f}={\mathfrak g}\oplus {\mathfrak p}\>,
\quad \text{with} \quad 
[{\mathfrak g},{\mathfrak g}]\subset {\mathfrak g}\>, 
\quad [{\mathfrak g},{\mathfrak p}]\subset 
{\mathfrak p}\>, 
\quad [{\mathfrak p},{\mathfrak p}]\subset {\mathfrak g}\>,
\label{CanonicalDec}
\end{equation}
where ${\mathfrak g}$ and $ {\mathfrak p}$ are the $+1$ and $-1$ eigenspaces of $\sigma_-$, respectively, and ${\mathfrak g}$ is the Lie algebra of $G$.
Then, the sigma model with target space a symmetric space can be described as a sigma model
with a field $f\in F$ where the $G$ action $f\to fg^{-1}$, $g\in G$, is
gauged. For instance, this is the approach described in the
prequel~\cite{Miramontes:2008wt} (see also \cite{Eichenherr:1981sk}). 
However, for present purposes,
we find it more convenient to work directly in the coset $F/G$ by
defining the $F$-valued field
\EQ{
{\cal F}=\sigma_-(f)f^{-1}
}
and working directly with ${\cal F}$ instead of $f$, which we
can think of as an $F$-valued field subject to the constraint
\EQ{
\sigma_-(\newf)=\newf^{-1}\ .
\label{cdo}
}
In this formalism there is no need to introduce gauge fields and this
simplification turns out to be useful. Of course, our approach in
terms of involutions can easily be translated into the gauged sigma
model language if need be (see Appendix~\ref{AppSigmaModel}).

We will also consider the principal chiral model which can either be
considered as a symmetric space $G\times G/G$ as above, with
$\sigma_-$ being the involution that exchanges the two $G$ factors,
or we can simply take the target space to be $F=G$ itself, in which
case the involution $\sigma_-$ is not required. We will take the
latter point of view in what follows.

The Lagrangian of the sigma model is simply
\EQ{
\LAG=-\frac1{8\kappa}\Tr{\cal J}_\mu  {\cal J}^\mu \ ,
}
where 
\EQ{
{\cal J}_\mu =\partial_\mu  \newf \newf^{-1}\ .
}
Note that $\newf\to \sigma_-(\newf)=\newf^{-1}$ is a symmetry of the action and
equations-of-motion and therefore it is consistent to impose it by
hand on the field $\newf$ from the start and, as mentioned above, 
in this formalism there are no gauge fields. 
The equations-of-motion for the group field are
\EQ{
\partial_\mu {\cal J}^\mu=0\ .
\label{p1}
}
In other words, $ {\cal J}_\mu$ provides the conserved currents corresponding to the
global $F_L\times F_R$ symmetry of the sigma model with target space $F$ (the
principal chiral model) under which $\newf\rightarrow U \newf V$ for any $U,V\in F$. The left and right currents are
\EQ{
{\cal J}^L_\mu=\partial_\mu \newf\newf^{-1}={\cal J}_\mu
\quad{\rm and}\quad
{\cal J}^R_\mu=\newf^{-1}\partial_\mu \newf 
={\cal F}^{-1}{\cal J}_\mu {\cal F},
}
and we can define the corresponding conserved charges
\EQ{
{\cal Q}_L=\int_{-\infty}^\infty dx\,\partial_0{\cal F}{\cal F}^{-1}\quad \text{and}\quad
{\cal Q}_R=\int_{-\infty}^\infty dx\,{\cal F}^{-1}\partial_0{\cal F}\>.
\label{cch}
}
In the principal chiral model these charges are independent. 
However, in the $F/G$ models the $F_L\times F_R$ symmetry is reduced by the constraint \eqref{cdo} so that they are invariant only under $\newf\to \sigma_-(U)\newf U^{-1}$ with $U\in F$. Taking~\eqref{NewField} into account, these transformations
correspond to $f\rightarrow U f$, which specifies the global symmetries
of the symmetric space sigma model in the gauged sigma model language. Consequently, in the $F/G$ models the two charges are related by $\sigma_-({\cal Q}_L)=-{\cal Q}_R$.

Since ${\cal J}_\pm=\partial_\pm \newf \newf^{-1}$, these currents trivially satisfy the Cartan-Maurer conditions\footnote{
In out notation $x_+=t + x$ and  $x_-=t- x$}
\EQ{
\partial_+{\cal J}_--\partial_-{\cal J}_+-[{\cal J}_+,{\cal J}_-]=0\ .
\label{p2}
}
Then, the equations-of-motion \eqref{p1}, along with the identity
\eqref{p2}, can be written in the form of a zero curvature condition:
\EQ{
\Big[\partial_+-\frac{{\cal J}_+}{1+\lambda},\partial_--\frac{{\cal J}_-}{1-\lambda}\Big]=0\ ,
\label{zcf}
}
where $\lambda$ is a spectral parameter. The residues at
$\lambda=\pm1$ then give the two equations
\EQ{
\mp\partial_\pm{\cal J}_\mp+\tfrac12[{\cal J}_+,{\cal J}_-]=0\ ,
\label{wee}
}
respectively, which are equivalent to \eqref{p1} and \eqref{p2}. 

\section{The Pohlmeyer Reduction}
\label{Pohl}

The Pohlmeyer reduction, at an algebraic level, 
involves imposing the conditions  (see
Appendix~\ref{AppSigmaModel}  and~\cite{Miramontes:2008wt})\footnote{In \cite{Miramontes:2008wt} the
  right-hand side had scales multipliers $\mu_\pm$. In the present
  work, we will not indicate these factors. We can either re-introduce
  them by scaling $x_\pm$, or one can think of them as having been
  absorbed into $\Lambda_\pm$ (see~Appendix~\ref{AppSigmaModel}).}
\EQ{
\partial_\pm\newf\newf^{-1}= f_\pm\Lambda_\pm f_\pm^{-1}\ ,
\label{laq}
}
where $\Lambda_\pm$ are constant elements in a maximal abelian subspace ${\mathfrak a}$ of ${\mathfrak p}$
in~\eqref{CanonicalDec} and $f_\pm\in F$. The natural degree-of-freedom left after the reduction is 
$\gamma=f_-^{-1}f_+$ which
is valued in $G\subset F$. In order to see this,
we act on \eqref{laq} with $\sigma_-$. The left-hand sides become
\EQ{
\sigma_-\big(\partial_\pm \newf\newf^{-1}\big)=
\partial_\pm \newf^{-1} \newf=-\newf^{-1}\partial_\pm \newf
}
while the right-hand sides transform into
\EQ{
\sigma_-\big(f_\pm\Lambda_\pm f_\pm^{-1}\big)=-\sigma_-(f_\pm)\Lambda_\pm
\sigma_-(f_\pm^{-1})\ .
}
The two can be made consistent by requiring
\EQ{
\sigma_-(f_\pm)=\newf^{-1}f_\pm\ ,
\label{jjj}
}
and so 
\EQ{
\sigma_-(\gamma)=f_-^{-1}\newf\newf^{-1}f_+=\gamma\ ,
\label{ValuedinG}
}
which shows that $\gamma\in G$. Actually, it is clear that $f_\pm$ are 
ambiguous
since we could always right-multiply by the group of elements that commute with
$\Lambda_\pm$, respectively. We shall
soon see that this freedom leads to a gauge symmetry in the reduced model.
Notice that once the reduction has been imposed the ``left'' $F$
charges can be written
\EQ{
{\cal Q}_L=\int_{-\infty}^\infty
dx\,\big( f_+\Lambda_+f_+^{-1}+ f_-\Lambda_-f_-^{-1}\big)\ .
}

Using \eqref{wee} with ${\cal J}_\pm=f_\pm\Lambda_\pm f_\pm^{-1}$, we have
\SP{
\big[f_+^{-1}\partial_-f_+ -\frac{1}{2}\gamma^{-1}\Lambda_-\gamma,\Lambda_+\big]=0\ .
}
This implies that
\EQ{
f_+^{-1}\partial_-f_+ -\frac{1}{2} \gamma^{-1}\Lambda_-\gamma=A_-^{(R)}
\label{laa}
}
where $A_-^{(R)}$ is an unknown element that 
satisfies $[A_-^{(R)},\Lambda_+]=0$ and, 
using~\eqref{jjj}, $\sigma_-\bigl(A_-^{(R)}\bigr)=A_-^{(R)}$. Therefore,
$A_-^{(R)}$ takes values in ${\mathfrak h}_+$, which is the Lie algebra
of the subgroup 
$H^{(+)}\subset G$ of elements that commute with $\Lambda_+$.
Similarly, we have
\EQ{
\big[-f_+^{-1}\partial_+f_++\gamma^{-1}\partial_+\gamma+\frac{1}{2}
\Lambda_+,\gamma^{-1}\Lambda_-\gamma\big]=0\ ,
}
which implies that
\EQ{
f_+^{-1}\partial_+f_+ -\gamma^{-1}\partial_+\gamma-\frac{1}{2}\Lambda_+=\gamma^{-1}A_+^{(L)}\gamma\
.
\label{lab}
}
Here, $[A_+^{(L)},\Lambda_-]=0$ and, using~\eqref{jjj} once more,
$\sigma_-\bigl(A_+^{(L)}\bigr)=A_+^{(L)}$. This shows that
$A_+^{(L)}\in{\mathfrak h}_-$, which is the Lie algebra
of the subgroup 
$H^{(-)}\subset G$ of elements that commute with $\Lambda_-$.

On the other hand, the integrability condition for \eqref{laq} implies
\EQ{
\big[\partial_+ - f_+\Lambda_+f_+^{-1},\partial_- - f_-\Lambda_-f_-^{-1}\big]=0\
,
}
from which we deduce
\EQ{
\big[\partial_++f_+^{-1}\partial_+f_+ - \Lambda_+,
\partial_-+f_+^{-1}\partial_-f_+ - \gamma^{-1}\Lambda_-\gamma\big]=0\
.
}
Using \eqref{laa} and \eqref{lab}, it gives
\EQ{
\big[\partial_++\gamma^{-1}\partial_+\gamma+\gamma^{-1}A_+^{(L)}\gamma-\frac{1}{2}\Lambda_+,
\partial_-+A_-^{(R)}-\frac{1}{2}\gamma^{-1}\Lambda_-\gamma\big]=0\>,
\label{www1}
}
which are the zero-curvature form of the 
Symmetric Space sine-Gordon (SSSG)
equations-of-motion. 
Notice that, as a consequence of~\eqref{laq}, this set of equations
has a natural $H_L^{(-)}\times
H_R^{(+)}$ gauge symmetry under which
\EQ{
f_\pm\longrightarrow f_\pm h_\pm^{-1}\ ,
}
where $h_\pm$ are local group elements in the subgroups
$H^{(\pm)}\subset G$. Under this symmetry
\EQ{
\gamma\longrightarrow h_- \gamma h_+^{-1}\ 
\label{gaugeLR1}
}
and 
\EQ{
A_-^{(R)}\longrightarrow h_+\big(A_-^{(R)}+\partial_-\big)h_+^{-1}\ ,\qquad
A_+^{(L)}\longrightarrow h_-\big(A_+^{(L)}+\partial_+\big)h_-^{-1}\ .
\label{gaugeLR2}
}
This is exactly the result of
\cite{Miramontes:2008wt}. 

The SSSG equations \eqref{www1} are integrable and lead to an infinite
set of conserved quantities which, as discussed in Appendix~\ref{AppConserved}, include charges corresponding to the global part
of the gauge group, and the energy and momentum. Since the conserved
charges play an important r\^ole, we will describe their construction
in some detail.
First of all, by projecting
\eqref{www1} onto ${\mathfrak h}_+$ and $\gamma\cdots\gamma^{-1}$ onto
${\mathfrak h}_-$ yields the zero curvature conditions
\EQ{
[\partial_++A_+^{(R/L)},\partial_-+A_-^{(R/L)}]=0\ ,
\label{Flatness}
}
where we have defined the ``missing'' components of the gauge connections, 
\SP{
A_+^{(R)}=&{\bf P}_{{\mathfrak h}_+}\big(
\gamma^{-1}\partial_+\gamma+\gamma^{-1}A_+^{(L)}\gamma\big)\ ,\\
A_-^{(L)}=&{\bf P}_{{\mathfrak h}_-}\big(
-\partial_-\gamma\gamma^{-1}+\gamma A_-^{(R)}\gamma^{-1}\big)\ .
\label{dmi}
}

Eq.~\eqref{Flatness} gives rise to the conserved quantities associated to the global version of the $H_L^{(-)}\times H_R^{(+)}$ gauge transformations.  Moreover, it enables the gauge fixing conditions that relate the SSSG equations to the non-abelian affine Toda equations ($A_\pm^{(R/L)}=A_\mp^{(R/L)}=0$), and the gauge fixing conditions required for their Lagrangian formulation (see~\eqref{yoo2}). Then, it is important to notice that~\eqref{Flatness} holds provided that
$\Lambda_\pm$ give rise to the orthogonal decompositions
\EQ{
{\mathfrak f} = \mathop{\rm Ker} \bigl( {\mathop{\rm Ad}}_{\Lambda_\pm}\bigr) \oplus \mathop{\rm Im} \bigl( {\mathop{\rm Ad}}_{\Lambda_\pm}\bigr)
\label{Orthogonal}
}
and, consequently, that
\SP{
&\bigl[\mathop{\rm Ker} \bigl( {\mathop{\rm Ad}}_{\Lambda_\pm}\bigr),
\mathop{\rm Ker} \bigl( {\mathop{\rm Ad}}_{\Lambda_\pm}\bigr)\bigr]
\subset
\mathop{\rm Ker} \bigl( {\mathop{\rm Ad}}_{\Lambda_\pm}\bigr)\>, \\[5pt]
&\bigl[\mathop{\rm Ker} \bigl( {\mathop{\rm Ad}}_{\Lambda_\pm}\bigr),
\mathop{\rm Im} \bigl( {\mathop{\rm Ad}}_{\Lambda_\pm}\bigr)\bigr]
\subset
\mathop{\rm Im} \bigl( {\mathop{\rm Ad}}_{\Lambda_\pm}\bigr)\>.
\label{Orthogonal2}
}
This is always true if the symmetric space $F/G$ is of definite signature ($G$ compact), which is the only case considered in this work. 
An example were the decomposition~\eqref{Orthogonal} is not satisfied is provided by the `` lightlike '' Pohlmeyer reduction of the sigma model with target space $AdS_n$ discussed in~\cite{Miramontes:2008wt}.

Under the gauge transformations~\eqref{gaugeLR1}--\eqref{gaugeLR2},
\EQ{
A_\pm^{(R)}\longrightarrow h_+\big(A_\pm^{(R)}+\partial_\pm\big)h_+^{-1}\ ,\qquad
A_\pm^{(L)}\longrightarrow h_-\big(A_\pm^{(L)}+\partial_\pm\big)h_-^{-1}\ .
\label{gaugeLR3}
}
Then, in order to construct gauge invariant conserved quantities, we will transform~\eqref{Flatness} into gauge invariant equations. First of all, we choose a gauge slice $\gamma_0$ such that any field $\gamma$ can be written as
\EQ{
\gamma=\phi_L\gamma_0\phi_R^{-1}\>,
\label{GaugeSliceSSSG}
}
with $\phi_L\in H^{(-)}$ and $\phi_R\in H^{(+)}$. Under gauge
transformations, $\gamma_0$ remains invariant while $\phi_L\to h_-\phi_L$ and
$\phi_R\to h_+\phi_R$. Then, it can be easily checked that
\EQ{
\tilde A_\pm^{(R)} = \phi_R^{-1}\bigl( A_\pm^{(R)}+\partial_\pm\big)\phi_R\quad{\rm and}\quad
\tilde A_\pm^{(L)} = \phi_L^{-1}\bigl( A_\pm^{(L)}+\partial_\pm\big)\phi_L
}
are gauge invariant and, moreover, that
\EQ{
[\partial_++\tilde A_+^{(R/L)},\partial_-+\tilde A_-^{(R/L)}]=0.
\label{FlatnessGI}
}

In general, $H^{(\pm)}$ will be
of the form $U(1)^{p_\pm}\times H_\text{ss}^{(\pm)}$, where $p_\pm$ are positive integers and $H_\text{ss}^{(\pm)}$
are semi-simple factors. This allows one to write
\EQ{
\phi_{R/L}= e^{\alpha_{R/L}} \varphi_{R/L}\>,
}
where $e^{\alpha_{R/L}}\in U(1)^{p_\pm}$ and $\varphi_{R/L}\in H_\text{ss}^{(\pm)}$.
Then, the projection of~\eqref{FlatnessGI} on the Lie algebras of $H_\text{ss}^{(\pm)}$ and $U(1)^{p_\pm}$ provide two different types of gauge invariant conserved quantities. 
Namely, the projection of~\eqref{FlatnessGI} on the Lie algebra of $U(1)^{p_\pm}$
shows that the currents
\EQ{
J^\mu_{R/L}=\epsilon^{\mu\nu}{\bf P}_{u(1)^{p_\pm}} \bigl(\tilde A_\nu^{(R/L)}\bigr)= \epsilon^{\mu\nu}\Bigl({\bf P}_{u(1)^{p_\pm}} \bigl(A_\nu^{(R/L)}\bigr)+ \partial_\nu \alpha_{R/L}\Bigr)
}
are conserved.
They lead to the ``local'' gauge invariant conserved quantities
\EQ{
Q_{R/L} = \int_{-\infty}^{+\infty} dx\> J^0_{R/L} =
\alpha_{R/L}(+\infty)- \alpha_{R/L}(-\infty) + 
\int_{-\infty}^{+\infty} dx\> {\bf P}_{u(1)^{p_\pm}} \bigl(A_1^{(R/L)}\bigr)
\label{LocalCharge}
}
which take values in the (abelian) Lie algebra of $U(1)^{p_\pm}$.
On the other hand, the projection of~\eqref{FlatnessGI} on the Lie algebra of $H^{(\pm)}_\text{ss}$  provide the ``non-local'' conserved quantities given by the path ordered exponentials
\SP{
\Omega_{R/L}&={\bf P}\exp\biggl( -\int_{-\infty}^\infty dx\>  {\bf P}_{{\mathfrak h}_\text{ss}^{(\pm)}} \Bigl(\tilde A_1^{(R/L)}\Bigr)\biggr)\\[5pt]
&=\varphi_{R/L}^{-1}(+\infty)\> {\bf P}\exp\biggl( -\int_{-\infty}^\infty dx\> {\bf P}_{{\mathfrak h}_\text{ss}^{(\pm)}}\Bigl(A_1^{(R/L)}\Bigr)\biggr)\>\varphi_{R/L}(-\infty)\ ,
\label{NonLocalCharge}
}
which take values in $H^{(\pm)}_\text{ss}$.
Notice that the conserved charges~\eqref{LocalCharge}
and~\eqref{NonLocalCharge} are {\it not\/} the same as the conserved
charges of the original sigma model ${\cal Q}_R$ and ${\cal Q}_L$. In
particular, the former are Lorentz invariant (see
Appendix~\ref{AppConserved}) while the latter are not. 
In certain
circumstances, and in particular for the soliton solutions, it can
transpire that 
for particular configurations ${\bf P}_{{\mathfrak h}_\text{ss}^{(\pm)}}\big(A_\pm^{(R/L)}\big)$ take values in an abelian
subalgebra of ${\mathfrak h}_\text{ss}^{(\pm)}$, and $\varphi_{R/L}$ in the corresponding abelian subgroup of $H_\text{ss}^{(\pm)}$ (for all $x$). In this case, the path
ordering in \eqref{NonLocalCharge} is unnecessary and we can 
write $\Omega_{R/L}=\exp Q^{(\text{ss})}_{R/L}$ for abelian charges $Q^{(\text{ss})}_{R/L}$
taking values in the relevant abelian subalgebras of ${\mathfrak h}_\text{ss}^{(\pm)}$. 

It is worth remarking that $\phi_L$ and $\phi_R$ are subject to
an ambiguity whenever a particular
field configuration is invariant under a certain subgroup of 
$H^{(-)}_L\times H^{(+)}_R$. As an example, consider the vacuum
configuration itself, $\gamma=1$ with $A_{\pm}^{(R/L)}=0$. It is
invariant under the global vector subgroup of $H^{(-)}_L\times
H^{(+)}_R$, which means that $\phi_L$ and $\phi_R$ are uniquely defined only modulo 
$\phi_L\to\phi_LU$ and $\phi_R\to\phi_RU$, with $U\in H^{(-)}_L\cap H^{(+)}_R$. Consequently, the local charges carried by the vacuum solution are unambiguously defined only up to
$Q_L\sim Q_L+\rho$ and $Q_R\sim Q_R+\rho$, for $\rho\in u(1)^{p_-}\cap u(1)^{p_+}$. So, in a sense, only the combination $Q_L-Q_R$ is an
unambiguously well-defined charge.
The significance of this and its relation to spontaneous
symmetry breaking will become clearer when we
discuss the Lagrangian formulation of the SSSG equations later in this
section.

The energy-momentum tensor is constructed 
in Appendix~\eqref{AppConserved}, and leads to the following
expression for the energy of a
configuration 
\SP{
{\mathscr E}=&\frac12\int dx\,\Tr\Big[-\bigl(\partial_+\gamma\gamma^{-1}+ A_+^{(L)}\bigr)^2+{A_+^{(R)}}^2\\
&-\bigl(\gamma^{-1}\partial_-\gamma- A_-^{(R)}\bigr)^2 +{A_-^{(L)}}^2+\Lambda_+\gamma^{-1}\Lambda_-\gamma-\Lambda_+\Lambda_-\Big]\ ,
\label{ener2}
}
relative to ${\mathscr E}=0$ for $\gamma=1$.
We will find that the dressing procedure always produces soliton
solutions of \eqref{www1} which have vanishing gauge fields
$A^{(R)}_-=A^{(L)}_+=0$ and which satisfy the conditions 
\EQ{
{\bf P}_{{\mathfrak
  h}_+}\big(\gamma^{-1}\partial_+\gamma\big)=0\ ,\qquad
{\bf P}_{{\mathfrak
    h}_-}\big(\partial_-\gamma\gamma^{-1}\big)=0\ ,
\label{Constraints}
}
and hence $A_\mu^{(R)}=A_\mu^{(L)}=0$. Then, the conserved charges only get
contributions from the boundary terms $\phi_{R/L}(\pm\infty)$ and, within the examples discussed in the following sections,
these turn out to be non-trivial only in the principal chiral models (see Section~\ref{PCM}).
As a consequence only the principal chiral model solitons are charged under the SSSG
$H_L^{(-)}\times H_R^{(+)}$ symmetry. In contrast, the solitons do always carry sigma model charge 
${\cal Q}_{L,R}$.
In Section~\ref{Beyond}, we will see how
to produce solitons in the reduced symmetric space sigma models 
which carry non-trivial $H_L^{(-)}\times H_R^{(+)}$ charges; 
however, one needs to go
beyond the dressing transformation to produce them.

\vspace{0.25cm}
\noindent
{\bf Lagrangian formulations}

It is only natural to search for a 
relativistically invariant Lagrangian
formulation of the SSSG equations \eqref{www1}. However, as we shall see and 
as has been pointed out
elsewhere \cite{Miramontes:2008wt,Bakas:1995bm,Grigoriev:2007bu} 
there are problems that arise in pursuing this idea,
and it may be that the SSSG equations
themselves should be used as a basis for a canonical quantization
without recourse to a Lagrangian.

Lagrangian formulations are only known when 
$H^{(-)}$ and $H^{(+)}$ are isomorphic 
and of the form~\cite{Miramontes:2008wt,Bakas:1995bm}
\EQ{
H^{(+)}_R=\epsilon_R(H)\ ,\qquad H^{(-)}_L=\epsilon_L(H)\ ,
}
where $H$ is a Lie group and $\epsilon_{L,R}:\,H\to G$ are two ``anomaly-free'' 
group homomorphisms that
descend to embeddings of the corresponding Lie algebras ${\mathfrak
  h}$ and ${\mathfrak g}$.\footnote{Here, anomaly free simply means that
$\Tr\bigl(\epsilon_L(a) \epsilon_L(b) \bigr)
=\Tr\bigl(\epsilon_R(a) \epsilon_R(b) \bigr)$ for all $a,b\in{\mathfrak h}$.
}
Then, each non-equivalent choice of $\epsilon_L$ and $\epsilon_R$
gives rise to a different Lagrangian formulation. This is obtained by
writing 
\EQ{
A_+^{(L)}=\epsilon_L({\cal A}_+)\ ,\qquad
A_-^{(R)}=\epsilon_R({\cal A}_-)\ ,
\label{yoo}
}
where ${\cal A}_\pm$ take values in ${\mathfrak h}$, and imposing the
constraints 
\SP{
&{\bf P}_{{\mathfrak h}_+}\Big(\gamma^{-1}\partial_+\gamma
+\gamma^{-1}\epsilon_L({\cal A}_+)\gamma\Big)=
\epsilon_R({\cal A}_+)\ ,\\[5pt]
&{\bf P}_{{\mathfrak h}_-}\Big(-\partial_-\gamma\gamma^{-1}
+\gamma\epsilon_R({\cal A}_-)\gamma^{-1}\Big)=
\epsilon_L({\cal A}_-)\>,
\label{gco2}
}
which can be viewed as a set of partial
gauge fixing
conditions~\cite{Grigoriev:2007bu,Miramontes:2008wt}.\footnote{In~\cite{Miramontes:2008wt}, it was shown that this interpretation is consistent provided that the orthogonal decompositions~\eqref{Orthogonal} hold, which is always true if the symmetric space is of definite signature.} They can be written as
\EQ{
A_-^{(L)}=\epsilon_L({\cal A}_-)\ ,\qquad
A_+^{(R)}=\epsilon_R({\cal A}_+)\ ,
\label{yoo2}
}
where $A_-^{(L)}$ and $A_+^{(R)}$ are the ``missing'' components defined in \eqref{dmi}.
These conditions 
reduce the $H_L^{(-)}\times H_R^{(+)}$ gauge symmetry~\eqref{gaugeLR1} to
\EQ{
\gamma\longrightarrow \epsilon_L(h)\gamma \epsilon_R(h^{-1})\>, \qquad h\in H\>,
\label{gaugeH}
}
under which ${\cal A}_\mu$ transforms as a gauge connection:
\EQ{
{\cal A}_\mu\longrightarrow h\big({\cal
  A}_\mu+\partial_\mu\big)h^{-1}\ .
\label{gaugeHA}
}
In addition, the gauge conditions~\eqref{gco2} leave a residual  symmetry under the global (abelian) transformations
\EQ{
\gamma\longrightarrow e^{\epsilon_L(\rho)}\gamma e^{+\epsilon_R(\rho)}\ ,\qquad 
{\cal A}_\mu\longrightarrow {\cal A}_\mu\>,
\label{GlobalH}
}
where $e^{\rho}$ is in the centre of $H$.

The gauge-fixed equations-of-motion are then
\EQ{
\big[\partial_++\gamma^{-1}\partial_+\gamma
+\gamma^{-1}\epsilon_L
({\cal A}_+)\gamma,\partial_-+\epsilon_R({\cal A}_-)\big]
=\frac{1}{4}[\Lambda_+,
\gamma^{-1}\Lambda_-\gamma]
\label{eom3}
}
and these follow as the 
equations-of-motion of the Lagrangian density
\SP{
\LAG&=\LAG_{WZW}(\gamma)+\frac1{2\pi}\Tr\Big(-\epsilon_L({\cal 
A}_+)\partial_-\gamma
\gamma^{-1}+\epsilon_R({\cal A}_-)\gamma^{-1}\partial_+\gamma\\
&+\gamma^{-1}\epsilon_L({\cal A}_+)\gamma\epsilon_R({\cal A}_-) -\epsilon_L({\cal A}_+)\epsilon_L({\cal A}_-)
-\frac{1}{4}\Lambda_+
\gamma^{-1}\Lambda_-\gamma\Big)\ ,
\label{ala}
}
where $\LAG_{WZW}(\gamma)$ is the usual WZW Lagrangian density 
for $\gamma$. In fact this theory is the 
asymmetrically gauged WZW model for $G/H$ specified by $\epsilon_{R/L}$ with a potential.
Notice that the partial gauge-fixing constraints \eqref{gco2} now appear as
the equations-of-motion of the gauge connection.
If we take the Lagrangian \eqref{ala} as the basis for a QFT 
then many questions arise. For instance are the resulting QFTs
independent of the choice of the form of the gauge group; {\it i.e.\/}, independent of $\epsilon_L$
and $\epsilon_R$? In many cases, it can be shown that different theories are
actually related by a target space T-duality symmetry~\cite{Miramontes:2004dr}, hinting
that they are equivalent at the quantum level. 

Now we turn to the symmetries of the Lagrangian theory and the
relation with the conserved charges $Q_L$ and $Q_R$ of the SSSG
equations. Since our primary interest is in the soliton solutions, it
is a fact that the transformations $\phi_{R/L}$ that bring $\gamma$ to the
gauge slice \eqref{GaugeSliceSSSG} lie in an abelian subgroup of
$H_L^{(-)}$ and $H_R^{(+)}$. As a consequence there are associated local conserved
 currents and charges.\footnote{For a more general configuration, we would have
 to separate out the abelian factors in $H=U(1)^p\times H_\text{ss}$ 
in an obvious way, 
as we did in the last section, and describe the semi-simple part in
terms of non-local conserved charges. However, for the soliton
solutions this technology is unnecessary.}
Then, for our purposes, it will be enough to restrict the following discussion to the case of abelian $H$.
Then, the Lagrangian~\eqref{ala} is symmetric under the (abelian) global transformations
\EQ{
\gamma \longrightarrow e^{\epsilon_L(u)} \gamma e^{-\epsilon_R(v)}\>,
\qquad {\cal A}_\mu\longrightarrow {\cal A}_\mu
\>, 
\label{GlobalPlus}
}
where $u,v$ take values in ${\mathfrak h}$. For $u=v$ this is just a global gauge transformation of the form~\eqref{gaugeH} while for $u=-v$ it is a global symmetry transformation of the form~\eqref{GlobalH}. Following standard means, 
we can derive the corresponding Noether currents as follows  (for instance, see~\cite{ZinnJustin:2002ru}).
Consider the variation of the Lagrangian action $S=\int d^2x\> \LAG$ under an infinitesimal transformation of the form
\EQ{
\gamma^{-1}\delta\gamma = \gamma^{-1}\epsilon_L(u) \gamma -\epsilon_R(v)
\>, \qquad \delta {\cal A}_\mu=0\>,
\label{GlobalPlusInf}
}
with $u=u(t,x)$ and $v=v(t,x)$. It reads
\SP{
\delta S&=\int d^2x\> \Tr\Big(\bigl[
\partial_++\gamma^{-1}\partial_+\gamma+\gamma^{-1}\epsilon_L({\cal A}_+)\gamma
-\frac{1}{2}\Lambda_+,\\
&
\hspace{6cm}
\partial_-+\epsilon_R({\cal A}_-)-\frac{1}{2}\gamma^{-1}\Lambda_-\gamma\big]\>\gamma^{-1}\delta\gamma\Bigr)\\[5pt]
&
=\int d^2x\> \Tr\bigg(\bigl( \partial_+{\cal A}_- -\partial_-{\cal A}_+\bigr) (u-v) + \\[5pt]
&
\hspace{3cm}
+\partial_- \Bigl( {\bf P}_{\mathfrak{h}_+} \bigl(\gamma^{-1}\partial_+\gamma +\gamma^{-1}\epsilon_L({\cal A}_+)\gamma\bigr) - \epsilon_R({\cal A}_+)\Bigr)\epsilon_R(v)+\\[5pt]
&
\hspace{3cm}
+\partial_+ \Bigl( {\bf P}_{\mathfrak{h}_-} \bigl(-\partial_-\gamma\gamma^{-1} +\gamma\epsilon_R({\cal A}_-)\gamma^{-1}\bigr) - \epsilon_L({\cal A}_-)\Bigr)\epsilon_L(u)\biggr)
}
Then, the condition that $\delta S$ vanishes for any $u,v$ provides the conservation equations we are looking for. Using the constraints~\eqref{gco2}, they read
\EQ{
\partial_+{\cal A}_- -\partial_-{\cal A}_+=0\>,
}
which are the conservation equations of the current
\EQ{
J^\mu =\epsilon^{\mu\nu} {\cal A}_\mu\>.
}
Notice that, since $\delta S=0$ for $u=v$, $J^\mu$ is the Noether current
associated to the abelian global transformations~\eqref{GlobalH}, and there is no conserved current associated to global gauge transformations.

$J^\mu$ is clearly not invariant
under the gauge transformations~\eqref{gaugeH}--\eqref{gaugeHA}, which in this (abelian) case are of the form
\EQ{
\gamma\longrightarrow e^{\epsilon_L(u)}\gamma e^{-\epsilon_R(u)}\>, \qquad 
{\cal A}_\mu\longrightarrow {\cal
  A}_\mu-\partial_\mu u\>.
\label{gaugeHAab}
}
In order to construct gauge invariant conserved quantities, we write the SSSG gauge slice~\eqref{GaugeSliceSSSG} as
\EQ{
\gamma=\phi_L\gamma_0\phi_R^{-1} =e^{\epsilon_L(\alpha+\beta)}\gamma_0 e^{-\epsilon_R(\alpha-\beta)}
\label{GaugeSliceLag}
}
such that, under~\eqref{gaugeHAab}, $\alpha\rightarrow \alpha+u$ while $\beta$ and $\gamma_0$ remain fixed. 
Then, the gauge invariant Noether current associated to the abelian global transformations~\eqref{GlobalH} is
\EQ{
\tilde J^\mu = \epsilon^{\mu\nu}\Bigl( {\cal A}_\nu + \partial_\nu \alpha\Bigr)\>,
}
which provides the Noether charge
\EQ{
Q^N= \alpha(+\infty) -\alpha(-\infty) + \int_{-\infty}^{+\infty} dx\> {\cal A}_1\>.
\label{NoetherCharge}
}
Similarly to the case of $Q_{R/L}$ discussed in the previous section, the definition of $Q^N$ is subject to an ambiguity whose form can be found by looking at the vacuum configuration $\gamma^{vac}=1$. Namely, since it is invariant under $\gamma\rightarrow e^\rho\gamma e^{-\rho}$, the field $\alpha$ in~\eqref{GaugeSliceLag} is only defined up to
$\alpha\rightarrow \alpha + \eta$ for any field $\eta\in {\mathfrak h}$ such that $\epsilon_L(\eta)=\epsilon_R(\eta)$. Consequently, 
the Noether charge is defined only modulo
\EQ{
Q^N \longrightarrow Q^N +q\quad \text{for each}\quad q\in{\mathfrak h}\quad \text{such that}\quad (\epsilon_L-\epsilon_R)(q)=0.
\label{NoetherChargeA}
}
In the Lagrangian formulation, this ambiguity has a physical interpretation. 
Notice that each constant $\rho\in{\mathfrak h}$ such that $(\epsilon_L-\epsilon_R)(\rho)=0$ generates a symmetry transformation of the form~\eqref{GlobalH} that changes $\gamma^{vac}=1$; namely,  $1\rightarrow e^{2\epsilon_L(\rho)}$.
Then, the ambiguity reflects the impossibility of defining a Noether charge for global symmetry transformations that do not leave the vacuum configuration invariant.

The relationship between the SSSG conserved quantities $Q_{R/L}$ (or $Q_{R/L}^{(\text{ss})}$) and $Q^N$ can be easily derived by taking into account~\eqref{LocalCharge},~\eqref{NonLocalCharge},~\eqref{yoo} and, according to~\eqref{GaugeSliceLag}, $\alpha_{R/L}=\epsilon_{R/L}(\alpha\mp \beta)$.
It reads
\EQ{
Q_L= \epsilon_L\bigl(Q^N+ Q^T\bigr) \>, \qquad
Q_R= \epsilon_R\bigl(Q^N- Q^T\bigr)\>,
}
where
\EQ{
Q^T = \beta(+\infty) -\beta(-\infty)
\label{TopologicalCharge}
}
is a kind of (gauge invariant) topological, or kink, charge. The definition of $Q^T$ only makes sense if the global symmetry~\eqref{GlobalH}, which corresponds to $\beta\rightarrow \beta+\rho$ in~\eqref{GaugeSliceLag}, changes the vacuum configuration and gives rise to non-trivial boundary conditions for $\gamma$. Consequently, the definition of $Q^T$ is also subject to an ambiguity whose form can be found by looking again at $\gamma^{vac}=1$. Since it is invariant under $\gamma\rightarrow e^\rho\gamma e^{-\rho}$, the field $\beta$ in~\eqref{GaugeSliceLag} is defined only up to $\beta\rightarrow \beta+\eta$ for any $\eta\in{\mathfrak h}$ such that $\epsilon_L(\eta)=-\epsilon_R(\eta)$, which means that the topological charge is defined modulo
\EQ{
Q^T\longrightarrow Q^T + q\quad \text{for each}\quad q\in{\mathfrak h}\quad \text{such that}\quad (\epsilon_L+\epsilon_R)(q)=0\>.
\label{TopologicalChargeA}
}

To summarize, in the Lagrangian formulation the soliton configurations are expected to carry both Noether $Q^N$ and topological $Q^T$ charges.
It is worth noticing that
the combination of the SSSG charges that is free of ambiguities reads
\EQ{
Q_L -Q_R = \bigl(\epsilon_L-\epsilon_R\bigr) (Q^N) + \bigl(\epsilon_L+\epsilon_R\bigr) (Q^T)
\label{QLQR}
}
which, not surprisingly, is also free of the ambiguities~\eqref{NoetherChargeA} and~\eqref{TopologicalChargeA}. Looking at this equation, it is worthwhile to recall that the different Lagrangian formulations of a set of SSSG equations are related by $H_L^{(-)}\times H_R^{(+)}$ gauge transformations, and that $Q_L -Q_R$ is gauge invariant and, hence, independent of the choice of $\epsilon_{R/L}$. Moreover, since T-duality transformations interchange Noether and topological charges, \eqref{QLQR} is consistent with the expectation that the different Lagrangian theories are indeed related by T-duality symmetries.

The physical meaning of the charges and their ambiguities
becomes clearer once we consider examples of particular gaugings.
The most obvious kind of gauging that can always be chosen is 
\EQ{
\epsilon_{L}(\alpha)=\epsilon_{R}(\alpha)=\alpha\ ,
}
which corresponds to gauging the vector subgroup of 
$H^{(-)}_L\times H^{(+)}_R$. 
Then, the value of $Q^N$ is meaningless, and the solitons are kinks characterized by the topological charge $Q^T$.
In this case, the Noether current corresponds to the axial transformations $\gamma\to e^\rho\gamma
e^\rho$, which do not leave the vacuum invariant. This means that at the classical level the symmetry is
spontaneously broken. Of course at the quantum level this would have
to be re-evaluated.

Since we are assuming that $H$ is abelian, one can also gauge the axial vector subgroup by taking 
\EQ{
\epsilon_{L}(\alpha)=-\epsilon_{R}(\alpha)=\alpha\ ,
}
In this case, $Q^N$ is free of ambiguities. It
is the Noether charge corresponding to vector
transformations $\gamma\to e^\beta\gamma e^{-\beta}$ that leave the vacuum invariant and, therefore, do not break the symmetry. In contrast, since the vacuum configuration is unique up to (axial) gauge transformations, the topological charge $Q^T$ is arbitrary. Therefore, solitons are similar to Q-balls. Other choices of $\epsilon_{R/L}$ give rise to different interpretations of solitons as some sort of dyons that carry both Noether and topological charge.

As we have mentioned, the dressing procedure always produces soliton
solutions of \eqref{www1} which have vanishing gauge fields
$A^{(R)}_-=A^{(L)}_+=0$. This means that the soliton solutions are
valid solutions of the gauged WZW model~\eqref{eom3}
and \eqref{gco2} for any choice of gauging with ${\cal
  A}_\mu=0$. Consequently, as is clear from~\eqref{NoetherCharge} and~\eqref{TopologicalCharge}, the charges can be
calculated in terms of 
boundary values of $\alpha$ and $\beta$ or, equivalently, $\alpha_{R/L}$.
The mass of the solitons can be calculated from the 
energy-momentum tensor of the gauged WZW theory which leads 
(up to an overall factor) to \eqref{ener2} with the identifications \eqref{yoo}.
Notice that, as a consequence of the anomaly free condition, $\Tr\bigl({A_\pm^{(R/L)}}^2 \bigr)=\Tr\bigl({A_\pm^{(L/R)}}^2 \bigr)=\Tr\bigl({\cal A}_\pm^2 \bigr)$.

\subsection{$\CP^2$ example}

In order to have an explicit example of the SSSG equations and their
Lagrangian formulation consider $\ms=\CP^2$. 
Since the symmetric space $\CP^2=SU(3)/U(2)$ has rank one, 
there is a unique Pohlmeyer reduction for which we can take (up to conjugation)
\EQ{
\Lambda_+=\Lambda_-\equiv\Lambda=\left( \begin{array}{ccc}
0 &-1 & 0 \\ 
1&0&0\\ 0&0&0
\end{array} \right)\ .
\label{cpp}
}
In this case
$H^{(\pm)}=U(1)$ and we can use both vector or axial gauging to achieve a
Lagrangian formulation. 
We can parameterize the group element and gauge field 
as (see~\eqref{GaugeSliceSSSG})
\EQ{
\gamma=e^{a_L h}\MAT{1&0&0\\ 0&\cos\theta e^{i\varphi}&\sin\theta\\
0&-\sin\theta &\cos\theta e^{-i\varphi}}
e^{-a_R h}\ ,
\label{ksa}
}
where $h=i\,\text{diag}(1,1,-2)$ is the generator of $\mathfrak h$, and $\alpha_{R/L}=a_{R/L} h$.

If we choose vector gauging, then solving the conditions \eqref{gco2} for
${\cal A}_\mu$
and writing $\alpha_L+\alpha_R= 2\alpha$ like in~\eqref{GaugeSliceLag}, yields the gauge invariant Noether 
current for axial transformations:
\EQ{
\tilde J^{(V)\mu}=\epsilon^{\mu\nu} \Bigl({\cal A}_\nu +\partial_\nu \alpha\Bigr)=
\frac13\Big(\big(\frac12+2\cot^2\theta)\partial^\mu(a_L-a_R)
+\cot^2\theta\partial^\mu\varphi\Big)h\ .
}
The two remaining equations, also depend only on $\alpha_L-\alpha_R$,
as one expects since the combination $\alpha_L+\alpha_R$ has been
gauged away, and
so it is convenient to define $a_L-a_R=\psi/2$ ($\psi h$ corresponds to $4\beta$ in~\eqref{GaugeSliceLag}):
\SP{
&\partial^\mu\partial_\mu\psi
=-4  \cos\theta\sin\varphi\ ,\\
&\partial^\mu\partial_\mu\theta+\frac{\cos\theta}{\sin^3\theta}\partial_\mu(
\varphi+\psi)\partial^\mu(\varphi+\psi)=-  \sin\theta\cos\varphi\ .
}
These equations along with the continuity of the Noether current follow
from the Lagrangian
\EQ{
\LAG=\partial_\mu\theta\partial^\mu\theta+\frac14\partial_\mu\psi
\partial^\mu\psi+\cot^2\theta\partial_\mu(\psi+\varphi)\partial^\mu(
\psi+\varphi)+2  \cos\theta\cos\varphi\ .
}
The Lagrangian manifests the axial symmetry $\psi\to\psi+a$. In this
case, the vacuum configuration is degenerate, $\gamma^{vac}=e^{h\psi/2}$,
{\it i.e.\/}~$\theta=\varphi=0$ with
$0\leq\psi<4\pi$, and the definition of the Noether charge does not make sense. Then, the solitons are characterized by the charge $Q^T$, which is simply the kink charge
\EQ{
Q^T=\frac14\big[\psi(\infty)-\psi(-\infty)\big]h\>.
}
Classically the axial symmetry would be spontaneously
broken. In the quantum theory, this would have to re-evaluated since
the theory is defined in $1+1$-dimensional spacetime, Goldstone's
Theorem does not apply and the would-be Goldstone
modes should be strongly coupled giving rise to a mass gap. 
A related issue is the fact 
that the Lagrangian does not have a good expansion in
terms of fields 
around their vacuum values due to the $\cot^2\theta$ term in the
Lagrangian. 

On the other hand, if we choose axial gauging, then $\alpha_L-\alpha_R=2\alpha$, and the gauge invariant conserved
Noether current for vector transformations is
\EQ{
\tilde J^{(A)\mu}=\epsilon^{\mu\nu} \Bigl({\cal A}_\nu +\partial_\nu \alpha\Bigr)=
\frac1{1+4\cot^2\theta}\Big(\frac 32\partial^\mu\big(a_L+a_R\big)
-\cot^2\theta\epsilon^{\mu\nu}\partial_\nu\varphi\Big)h\ .
}
Then, we can define $a_L+a_R=\tilde\psi/2$ ($\tilde\psi h=4\beta$ in~\eqref{GaugeSliceLag}) and the
corresponding Lagrangian is
\SP{
\LAG=\partial_\mu\theta\partial^\mu\theta+\frac1{1+4\cot^2\theta}&
\Big(\frac94\partial_\mu\tilde\psi
\partial^\mu\tilde\psi+\cot^2\theta\partial_\mu\varphi\partial^\mu
\varphi\\ &
-6\cot^2\theta\epsilon^{\mu\nu}\partial_\mu\tilde\psi\partial_\nu
\varphi\Big)+2  \cos\theta\cos\varphi\ ,
\label{axiallag}
}
which manifests the vector symmetry $\tilde\psi\to\tilde\psi+a$. In
this case the vacuum is non-degenerate, $\gamma^{vac}=1$, 
because $\tilde\psi$ is not a
good coordinate around $\theta=\varphi=0$,\footnote{In the same way
that the polar angle it not a good coordinate around $r=0$.} 
and the vacuum
is invariant under the (vector) symmetry. Consequently, the definition of $Q^T$ does not make sense. In contrast, 
\EQ{
Q^N=\int_{-\infty}^{+\infty} dx\> \tilde J^{(A)}_0
}
is unambiguously defined. Moreover, in this case the Lagrangian
does have a  good field expansion around the vacuum.

\section{Giant Magnons and their Solitonic Avatars}
\label{GM}

``Giant magnon'' is the name given to a soliton of the reduced $F/G$ model
in the context of string theory. In particular, the examples of $S^5$
and $\CP^3$ are directly relevant to the AdS/CFT correspondence for
$AdS_5\times S^5$~\cite{Maldacena:1997re} and $AdS_4\times\CP^3$~\cite{NewDuality}, respectively. In
this section, we review the known giant magnons in the context of 
$S^n$ and $\CP^n$ and use them to illustrate some of the more important ideas discussed in the previous
section in a more concrete way. In particular, we will discuss the
relation between the giant magnons and their 
relativistic solitonic avatars in the associated SSSG equations.

The sphere $S^n$ corresponds to the symmetric space
$SO(n+1)/SO(n)$ while the complex symmetric space $\CP^n$ 
corresponds to $SU(n+1)/U(n)$. In both cases the
associated involution is
\EQ{
\sigma_-(\newf)=\theta \newf\theta^{-1}\ ,
\label{iv3}
}
where 
\EQ{
\theta=
\text{diag}\big(-1,1,\ldots,1\big)\ .
\label{xx2}
}

In Appendix \ref{Appscp} we explain how to map the spaces $S^n$ and
$\CP^n$, expressed in terms of their usual coordinates, into the group field
${\cal F}$. For the spheres, parameterized by a real unit $n+1$-vector $\BX$
with components $X_a$, $|\BX|=1$, we have 
\EQ{
\newf=\theta\left(1-2\BX\BX^T\right)\ ,
\label{gqq2}
}
while for the complex projective spaces $\CP^n$ 
we have the complex $n+1$ vector $\BZ$ whose components are 
the complex projective
coordinates $Z_a$, $a=1,\ldots,n+1$, 
so that $\CP^n$ is identified by modding out by complex re-scalings
$Z_a\sim\lambda Z_a$, $\lambda\in\C$. In this case
\EQ{
\newf=\theta\left(1-2\frac{\BZ\BZ^\dagger}{|\BZ|^2}\right)\ .
\label{gqq1}
}

The giant
magnons can be though of as excitations around a ``vacuum'' which is
the simplest solution to the equations-of-motion and Pohlmeyer
constraints, \eqref{p1} and \eqref{laq}. The vacuum solution has $f_\pm=1$ and
\EQ{
\newf_0=\exp\big[ x_+\Lambda_++  x_-\Lambda_-\big]\ ,
\label{yhh}
}
where, up to overall conjugation and generalizing \eqref{cpp},
\EQ{
\Lambda_+=\Lambda_-\equiv\Lambda=\left( \begin{array}{cc|c}
0 &-1 & 0 \\ 
1&0&0\\ \hline 0&0&0
\end{array} \right)\ .
\label{cpp2}
}
This solution corresponds to
the following solution for the sphere and projective coordinates,
\EQ{
\BX_0=\BZ_0=\ee_1\cos t-\ee_2\sin t\ .
}
Here, $\ee_1,\ldots,\ee_{n+1}$ are a set of orthonormal vectors in
$\R^{n+1}$. 
The physical interpretation is clear: the string is collapsed to a
point which traverses a great circle on $S^n$ defined by the plane
spanned by $\ee_1$ and $\ee_2$ at the speed of 
light.\footnote{Notice that the
  plane is determined by the choice of representative $\Lambda$.}
The vacuum solution actually carries infinite sigma model charge and
so the physically meaningful charge is actually the charge relative to
the vacuum,
\EQ{
\Delta{\cal Q}_L=\int_{-\infty}^\infty dx\,\big(
\partial_0{\cal F}{\cal F}^{-1}-
\partial_0{\cal F}_0{\cal F}_0^{-1}\big)\ .
\label{cch2}
}
In particular, the component of this charge along the Lie algebra
element $\Lambda$, up to scaling, 
is identified with $\Delta-J$, the difference between the scaling
dimension and $R$ charge of the associated operator in the boundary CFT:
\EQ{
\Delta-J=\frac{\sqrt\lambda}{8\pi}\Tr\,\big(\Lambda \Delta{\cal Q}_L\big)\ ,
}
where $\lambda$ is the 't~Hooft coupling.

For the sphere case, we can express the conserved charge \eqref{cch2}
directly in terms of $X$:
\EQ{
{\cal Q}_{L,ab}=\int_{-\infty}^\infty dx\,\big(\partial_0X_aX_b
-X_a\partial_0X_b\big)\ .
\label{cch3}
}
In particular, $\Delta-J=\tfrac{\sqrt\lambda}{2\pi}
\Delta{\cal Q}_{L,12}$.
There is an analogous equation for the complex projective
spaces in terms of the projective coordinates.

\vspace{0.25cm}
\noindent
{\bf $\boldsymbol{S^n}$ giant magnons}

The original giant magnon was described by 
Hofman and Maldacena \cite{Hofman:2006xt}. 
It is a solution which takes values in the
subspace $S^2\subset S^n$ picked out by 3 mutually orthonormal vectors
$\{\ee_1,\ee_2,\BOmega\}$. 
The vectors $\ee_1$ and $\ee_2$ are already fixed by the
choice of vacuum solution, however the direction of $\BOmega$, which  
describes an $S^{n-2}\subset S^n$, plays the r\^ole of an
internal collective coordinate of the magnon.
If $(\theta,\phi)$ are polar coordinates on $S^2$, then the solution
written down by Hofman and Maldacena is
\SP{
\cos\theta&=\frac{\sin\tfrac p2}{\cosh x'}\ ,\\
\tan(\phi-t)&=\tan\frac p2\tanh x'\ .
\label{hmm}
}
Here, and in the following, we define the Lorentz boosted coordinates
$t'$ and $x'$:
\EQ{
x'=x\cosh\vartheta-t\sinh\vartheta\ ,\qquad
t'=t\cosh\vartheta-x\sinh\vartheta\ , 
\label{boost}
}
where $\vartheta$ is the rapidity ($v=\tanh\vartheta$). For the
Hofman-Maldacena magnon,
\EQ{
\tanh\vartheta=\cos\frac p2\ .
}
Notice that the magnon is not relativistic in the sense that
the moving solution {\it is not} the Lorentz boost of the stationary solution.
The reason is that the Pohlmeyer constraints \eqref{laq} 
are not Lorentz covariant (this is discussed in more detail in
Appendix \ref{AppConserved}).
In terms of the unit vector $\BX$, we can write this solution as
\SP{
\BX=&\big[
\sin t\sin\tfrac p2\tanh x'-\cos t\cos\tfrac p2\big]\ee_1\\ &+
\big[\cos t\sin\tfrac p2\tanh
x'+\sin t\cos\tfrac p2\big]\ee_2+\sin\tfrac p2\sech x'\BOmega\ .
\label{hmmx}
}
It has sigma model charge 
\EQ{
\Delta{\cal Q}_L=-4\big|\sin\frac p2\big|\,\Lambda\ ,
}
relative to the vacuum. 

The solitonic avatar of the Hofman-Maldacena giant magnon 
in the reduced SSSG model has vanishing 
gauge fields $A_+^{(L)}=A_-^{(R)}=0$, while
the non-vanishing elements of $\gamma$ are 
\EQ{
\gamma=\left(\begin{array}{cc|c}-1 & 0 & \B0^T\\ 
0 &-\cos\theta(x) & \sin\theta(x)\,\BOmega^T \\ \hline
\B0 & \sin\theta(x)\,\BOmega &
{\bf1}+(\cos\theta(x)-1)\,\BOmega\BOmega^T\end{array}\right)\ ,
\label{sal}
}
where we have highlighted the $2\times2$ subspace associated to
$\ee_1$ and $\ee_2$. In the above, 
$\theta(x)$ (not to be confused with the polar angle above or
the rapidity) 
is the soliton solution to the sine-Gordon equation 
\EQ{
\partial_\mu\partial^\mu\theta=-\sin\theta\ ,
}
which can be written 
\EQ{
\theta=4\tan^{-1}(e^x)\ .
}
Since the sine-Gordon equation 
is relativistic, in the sense that the moving solution is the Lorentz
boost of the static solution, it is sufficient to write the solution above in 
the soliton rest frame. 
This soliton has vanishing charges $Q_L=Q_R=0$.~\footnote{Since $H=SO(n-1)$ is semi-simple for $n\geq4$, these charges provide examples of the conserved quantities $Q^{(\text{ss})}_{R/L}$ defined in the paragraph after~\eqref{NonLocalCharge}.}

The second kind of solution is 
Dorey's dyonic giant magnon \cite{DyonicGM}. 
The relation of Dorey's magnon to the Hofman-Maldacena magnon is
analogous to the relation between the dyon and monopole solutions 
in gauge theories in $3+1$ dimensions. 
In the latter case, the dyon is obtained by allowing the
charge angle,
a collective coordinate talking values in $S^1$, 
to move around the circle with constant velocity. The
non-trivial aspect of this is that the angular motion has a
back-reaction on the original monopole. One way to think of what is
happening is in terms of Manton's picture of geodesic motion
\cite{Manton:1981mp}. 
The charge angle is
an internal collective coordinate of the monopole and the idea is that
one can make a time-dependent solution by allowing the internal 
collective coordinates to be time dependent. For low velocities the
motion is simply geodesic motion on the moduli space defined by a
metric which is constructed from the inner-product of the zero modes
associated to the collective coordinates. In the present setting, 
it is not clear whether Manton's analysis applies
  directly because the Hofman-Maldacena giant magnon is a
  time-dependent solution rather than a time-independent one like the 
monopole. We have seen that the Hofman-Maldacena giant magnon has 
an internal collective coordinate
$\BOmega$ which parameterizes an $S^{n-2}$. Dorey's solution
corresponds to allowing $\BOmega$ to move around a great circle 
(the geodesic) in
$S^{n-2}$. We can describe this motion by picking out two 
orthonormal vectors $\BOmega^{(i)}$, orthogonal to $\ee_1$ and $\ee_2$, 
and then take (in the magnon's rest frame)
\EQ{
\BOmega(t)=\cos(t\sin\alpha)\BOmega^{(1)}+
\sin(t\sin\alpha)\BOmega^{(2)}\ .
\label{nge1}
}
The parameter $\alpha$ sets the angular velocity.
This motion has a back-reaction on the original solution and we can write
the complete moving solution as
\SP{
\BX=&\big(-\cos t\cos\tfrac p2+
\sin t\sin\tfrac p2\tanh(x'\cos\alpha)\big)\ee_1\\ &+
\big(\sin t\cos\tfrac p2+
\cos t\sin\tfrac p2\tanh(
x'\cos\alpha)\big)\ee_2+\sin\tfrac p2\sech(x'\cos\alpha)\BOmega(t')\ ,
\label{dm4}
}
The parameter $\alpha$ and the rapidity $\vartheta$ are determined by
two parameters $p$ and $r$ via
\EQ{
\cot\alpha=\frac{2r}{1-r^2}\sin\frac p2\ ,\qquad
\tanh\vartheta=\frac{2r}{1+r^2}\cos\frac p2\ .
\label{defa}
}
The Hofman-Maldacena magnon corresponds to the limit $r\to1$ (or $\alpha\rightarrow0$).

Notice that the back-reaction of the angular motion is simply taken
care of by the 
replacement $x'\to x'\cos\alpha$. The dyonic giant magnon carries charge
\EQ{
{\cal Q}_L=-\frac{2(1+r^2)}{r}\big|\sin\frac p2\big|\,\Lambda
-\frac{2(1-r^2)}{r}\big|\sin\frac p2\big|\, h\ ,
}
relative to the vacuum,
where $h$ is the generator of $SO(n+1)$ corresponding to rotations in
the plane picked out by $\BOmega^{(i)}$:
\EQ{
h=\BOmega^{(1)}\BOmega^{(2)T}-
\BOmega^{(2)}\BOmega^{(1)T}\ .
\label{sq1}
}

In the reduced SSSG model, the dyonic magnon gives a soliton 
for which the gauge fields do not vanish:
\EQ{
A_+^{(L)}=-A_-^{(R)}=\frac{\cos^2\alpha\sin\alpha}{\cos(2\alpha)-
\cosh(2x\cos\alpha)}h\ .
\label{zii}
}
In addition, the ``missing'' components defined in \eqref{dmi} are
\EQ{
A_+^{(R)}=-A_-^{(R)}\ ,\qquad A_-^{(L)}=-A_+^{(L)}\ .
}
which means that only the temporal components of the currents
$J^\mu_{R/L}$ are non-vanishing. In addition, we have
$A_\mu^{(L)}=-A_\mu^{(R)}$ which, in the case when $H$ is abelian,
corresponds to the condition for axial gauging \eqref{yoo} in the
Lagrangian formulation.
The group field (in the rest frame) generalizes \eqref{sal} in an obvious way:
\EQ{
\gamma=\left(\begin{array}{cc|c}-1 & 0 & \B0^T\\ 
0&-\cos\theta(x)& \sin\theta(x)\,\BOmega(t)^T\\ \hline
\B0&\sin\theta(x)\,\BOmega(t) &{\bf1}+
(\cos\theta(x)-1)\BOmega(t)\BOmega(t)^T
\end{array}\right)\ .
\label{sal2}
}
where
\EQ{
\cos\theta(x)=1-2\cos^2\alpha\sech^2(x\cos\alpha)\ ,
}
which includes the effects of the back reaction of the geodesic motion.
The solution carries charges \eqref{LocalCharge} 
\EQ{
Q_L=-Q_R=\int_{-\infty}^\infty dx\,\frac{\cos^2\alpha\sin\alpha}
{\cos(2\alpha)-\cosh(x\cos\alpha)}h=
\Big(\alpha-\frac\pi2\Big)h\ .
}
In particular, notice that the non-vanishing combination $Q_L-Q_R$ is
the unambiguously defined charge according to the discussion in
Section \ref{Pohl}.
Notice that for these solutions there is no contribution from the boundary
terms in  \eqref{LocalCharge}. In addition, 
when $\ms=S^n$, $n>3$, the subgroup $H=SO(n-1)$ is
non-abelian. However, we can still define local conserved currents and
associated charges because $\phi_{R/L}$ and gauge fields $A_\mu^{(R/L)}$ 
lie in an abelian subgroup $SO(2)\subset H$. The dyonic soliton has a mass
\EQ{
M=4\cos\alpha\ .
}
The Lagrangian interpretation of these dyons depends of the choice of $\epsilon_{R/L}$, which fixes the form of the group of gauge transformations.
In the gauged WZW Lagrangian formulation with vector gauging, which can be achieved for any~$n$, these dyons would
carry non-vanishing topological charge $Q^T$. 
However, for the particular case of
$\ms=S^3$, when $H=SO(2)$ is abelian, it is also possible to define 
an axially gauged WZW theory, 
in which case the dyons carry non-vanishing Noether charge 
$Q^N$ corresponding to vector $SO(2)$ transformations. In both cases, $Q^T$ and $Q^N$ are related to $Q_{R/L}$ by means of~\eqref{QLQR}.

\vspace{0.25cm}
\noindent
{\bf $\boldsymbol{\CP^n}$ giant magnons}

Motivated by its application to the investigation of the AdS/CFT correspondence for $AdS_4\times \CP^3$~\cite{NewDuality}, the $\CP^3$ case has been discussed in some detail in the literature~\cite{Gaiotto:2008cg,Grignani:2008is,Abbott:2008qd,Astolfi:2008ji,Grignani:2008te}.
The giant magnon solutions described so far are all obtained by
embeddings of the Hofman-Maldacena giant magnon and Dorey's dyonic magnon. 
The Hofman-Maldacena giant magnon can be embedded in $\CP^n$ in two
distinct ways. Firstly, by taking $S^2\simeq \CP^1\subset\CP^n$
\cite{Gaiotto:2008cg}.
If $\theta(x,t)$ and $\phi(x,t)$ is the solution in terms of 
polar coordinates in \eqref{hmm}, then
the projective coordinates are
\EQ{
\BZ=e^{i\phi(2x,2t)/2}\sin\big(\theta(2x,2t)/2\big)\ee_1
+e^{-i\phi(2x,2t)/2}\cos\big(\theta(2x,2t)/2\big)\ee_2\ .
\label{ma9}
}
The scaling of the spacetime coordinates here is necessary in order to be
consistent with the scaling of the Pohlmeyer constraints in 
\eqref{laq}. In addition, in
order that the solution is oriented with respect to the choice of
vacuum in \eqref{yhh}, we have to rotate it with an element of
$SU(2)\subset SU(n)$, $\BZ\to U\BZ$, 
\EQ{
U=\frac1{\sqrt{2}}\left(\begin{array}{cc|c}
e^{\pi i/4}&e^{i\pi/4}& 0\\ e^{3i\pi/4}&e^{-i\pi/4}& 0\\ \hline 
0&0&1\end{array}\right)\ .
}
Notice that this solution has no internal collective coordinates.
The charge carried by the magnon is
\EQ{
\Delta{\cal Q}_L=-2\big|\sin\frac p2\big|\,\Lambda\ ,
}
relative to the vacuum.

This magnon corresponds to a soliton solution of the SSSG equations of the form
\EQ{
\gamma=\left(\begin{array}{cc|c}e^{i\psi} &0&0\\ 0&e^{-i\psi}&0\\ \hline
0&0&1\end{array}\right)\ ,
}
with $A_+^{(L)}=A_-^{(R)}=0$. The
field $\psi$ then satisfies the sine-Gordon equation
\EQ{
\partial^\mu\partial_\mu\psi=2\sin(2\psi)\ .
}
The giant magnon solution \eqref{ma9} corresponds to the sine-Gordon kink
\EQ{
\psi=2\tan^{-1}(e^{2x})+\frac\pi2\ .
}
The solution carries $H_L^{(-)}\times H_R^{(+)}$
charge 
\EQ{
Q_L=-Q_R=\frac12\big(\psi(\infty)-\psi(-\infty)\big)=\frac\pi2\ .
}

The inequivalent embedding of the
Hoffman-Maldacena giant magnon is via the subspace
$\RP^2\subset\CP^n$ \cite{Grignani:2008is}. 
The solution for $\BZ$ is exactly equal to $\BX$ in
\eqref{hmmx}, however the vector $\BOmega$ can now be taken to be a
  complex vector with $|\BOmega|=1$.\footnote{In addition, we
  should replace $\BOmega^T$ by $\BOmega^\dagger$ in~\eqref{sal}.} Consequently the soliton has internal collective
coordinates associated to an $S^{2n-3}$. 
The fact that the solution is valued in
$\RP^2$ is because the solution \eqref{hmmx} is itself valued in
$S^2$ and this fixes the 
complex scaling freedom, $\BZ\to\lambda\BZ$, up to the discrete element
$\BZ\to-\BZ$ and a further quotient by this gives $\RP^2$.
The dyonic magnon can be embedded in an analogous way when $n\geq3$ 
via the subspace $\RP^3\subset\CP^n$ \cite{Abbott:2008qd}.

\section{Magnons and Solitons by Dressing the Vacuum}
\label{dress}

One way to construct the magnon/soliton solutions is to use an approach 
known as the 
dressing transformation \cite{Dressing} 
which is closely related to the B\"acklund transformation. 
For magnons in
string theory this approach has been described in detail in 
\cite{Spradlin:2006wk,Jevicki:2007pk}. For the SSSG theories, and
in particular their gauged WZW formulations, such an
approach was described in \cite{Bakas:1995bm,Park:1994bx}.  
Schematically the transformation takes a 
known solution---for example a trivial kind of 
solution that we call the ``vacuum''---and adds in a
soliton.\footnote{In Section~\ref{PCM}, we will show that in the principal chiral model cases multiple soliton solutions are sometimes produced where the solitons are all mutually at rest.}
It is an important fact that the dressing transformation is consistent
with the Pohlmeyer reduction in the sense 
that if the original solution satisfies the constraint \eqref{laq}
then so will the dressed solution. In fact we shall see that the
dressing transformation constructs both the magnon and its SSSG
soliton avatar at the same time without the need to map one into the other.

The dressing procedure has been described for general symmetric space
sigma models in \cite{Harnad:1983we} and we shall draw
heavily on results derived there. The procedure begins by identifying a 
``vacuum'' solution. In the present context, our vacuum solution will
be the simplest solution that satisfies the Pohlmeyer constraint
\eqref{laq}. This identifies it as \eqref{yhh}
which naturally satisfies the Pohlmeyer constraints
\eqref{laq} with $f_\pm=1$. In Appendix~\ref{AppSSSGEoM}, we show that
the dressing dressing transformation directly produces a soliton of
the SSSG equations \eqref{www1} 
with vanishing $A_+^{(L)}=A_-^{(R)}=0$:  
\EQ{
\partial_-\big(\gamma^{-1}\partial_+\gamma\big)=\frac{1}
4[\Lambda_+,\gamma^{-1}\Lambda_-
\gamma]\ .
\label{eom}
}
In addition, we find that the conditions 
\EQ{
\gamma^{-1}\partial_+\gamma\Big|_{{\mathfrak
    h}_+}=\partial_-\gamma\gamma^{-1}\Big|_{{\mathfrak h}_-}=0
\label{gco}
}
are satisfied.\footnote{This latter result makes use of the orthogonal decompositions~\eqref{Orthogonal}, which hold in general for symmetric spaces of definite signature.}
Notice that this means that the solitons automatically
satisfy the equations of the gauged WZW model \eqref{ala} for any
choice of gauging.

The strategy of \cite{Harnad:1983we} begins by defining the symmetric space
sigma model $F/G$ in terms of initially the subgroup
$F\subset SL(n,\C)$ via one, or possibly more, involutions that we
denote collectively as $\sigma_+$. In order to pick out the coset
$F/G\subset F$, the second part of the construction involves
the extra involution $\sigma_-$ whose explicit form for $S^n$ and $\CP^n$ have been given in \eqref{iv3}.
The dressing transformation is then constructed in 
$SL(n,\C)$ and the involutions give constraints
that ensure that the transformation is restricted to the quotient $F/G$.
In this work, we shall focus on three examples in order to be concrete:

(i) $F/G=SU(n)\times SU(n)/SU(n)$. These are the principal chiral
models and in this case it is more convenient to 
formulate the sigma model directly in terms of an
$SU(n)$-valued field $\newf(x)$. In this case, there is only a single
involution
\EQ{
\sigma_+(\newf)={\newf^\dagger}^{-1}
\label{inv1}
}
required, and the involution $\sigma_-$ is absent.

(ii) The complex projective spaces 
$\CP^n=SU(n+1)/U(n)$. In this case, there
are two involutions 
$\sigma_+(\newf)={\newf^\dagger}^{-1}$ and $\sigma_-(\newf)=\theta 
\newf\theta$, the latter defined in \eqref{iv3}.

(iii)  $S^n=SO(n+1)/SO(n)$. In this case, there
are three involutions 
$\sigma_+^{(1)}(\newf)={\newf^\dagger}^{-1}$, $\sigma_+^{(2)}(\newf)=\newf^*$ 
and $\sigma_-(\newf)=\theta \newf\theta$, the latter defined in \eqref{iv3}.

Starting in $SL(n,\C)$, 
the equations-of-motion for the sigma model have the zero curvature form
\eqref{zcf} which are the integrability conditions for the associated linear system
\SP{
\partial_+\Psi(x;\lambda)&=\frac{\partial_+\newf\newf^{-1}}
{1+\lambda}\Psi(x;\lambda)\ ,\\
\partial_-\Psi(x;\lambda)&=\frac{\partial_-\newf\newf^{-1}}
{1-\lambda}\Psi(x;\lambda)\ .
\label{hqq}
}
Notice that the group field is simply
\EQ{
\newf(x)=\Psi(x;0)\ .
\label{zlz}
}

The dressing transformation involves constructing 
a new solution $\Psi$ of the linear system of the form 
\EQ{
\Psi(x;\lambda)=\chi(x;\lambda)\Psi_0(x;\lambda)
\label{drs}
}
in terms of an old one
$\Psi_0$, which in our case corresponds to the vacuum solution in \eqref{yhh}:
\EQ{
\Psi_0(x;\lambda)=\exp\Big[\frac{ x_+}{1+\lambda}\Lambda_++\frac{ x_-}
{1-\lambda}\Lambda_-\Big]\ ,
\label{vso}
}
By picking out the residues of
$\partial_\pm\Psi(\lambda)\Psi(\lambda)^{-1}$ at $\lambda=\mp1$ (which
come entirely from the terms where the derivatives hit
$\Psi_0(x;\lambda)$) it follows that
\EQ{
\partial_\pm \newf \newf^{-1}= \chi(\mp1)\Lambda_\pm\chi(\mp1)^{-1}\ . 
\label{oss}
}
This is a key result because it means that the dressing
transformation preserves the Pohlmeyer reduction and, in addition, we
have
\EQ{
f_\pm=\chi(\mp1)\Phi\ ,
}
where $\Phi$ is a, as yet, unknown element that commutes with
$\Lambda_\pm$:
\EQ{
[\Lambda_\pm,\Phi]=0\ ,
}
which will be chosen so that 
\EQ{
\gamma=f_-^{-1}f_+=\Phi^{-1}\chi(+1)^{-1}\chi(-1)\Phi
\label{xxd}
}
is valued in $G\subset F$.
This will then guarantee that
the dressing transformation will give a soliton solution in the
associated SSSG model. In addition,
we will see that $f_\pm$ satisfy \eqref{jjj}. We will find below that
\EQ{
\Phi=\newf_0^{1/2}=\exp\Big[\frac{ x_+\Lambda_+}2+ \frac{ x_-\Lambda_-}2\Big] .
\label{dph}
}

We now briefly review the construction of the dressing factor $\chi(\lambda)$
following \cite{Harnad:1983we}. The general form is
\EQ{
\chi(\lambda)=1+\sum_i\frac{Q_i}{\lambda-\lambda_i}\ ,\qquad
\chi(\lambda)^{-1}=1+\sum_i\frac{R_i}{\lambda-\mu_i}
}
where the residues $Q_i$ and $R_i$ are matrices of the form
\EQ{
Q_i=\BX_i\BF_i^\dagger\ ,\qquad R_i=\BH_i\BK_i^\dagger\ ,
\label{ResiduesMatrices}
}
for vectors $\BX_i$, $\BF_i$, $\BH_i$ and $\BK_i$.\footnote{Notice that $i$ is not the
vector index but rather labels a set of vectors associated to the poles of $\chi^{\pm1}(\lambda)$.}

Taking the residues of $\chi(\lambda)\chi(\lambda)^{-1}=1$ at
$\lambda=\lambda_i$ and $\mu_i$, gives, respectively, 
\EQ{
Q_i+\frac{Q_iR_j}{\lambda_i-\mu_j}=0\ ,\qquad
R_i+\frac{Q_jR_i}{\mu_i-\lambda_j}=0\ ,
\label{xzl}
}
which can be used to solve for $\BX_i$ and $\BK_i$:
\EQ{
\BX_i\Gamma_{ij}=\BH_j\ ,\qquad \BK_i(\Gamma^\dagger)_{ij}=-\BF_j\ ,
}
where the matrix
\EQ{
\Gamma_{ij}=\frac{\BF_i^\dagger \BH_j}{\lambda_i-\mu_j}\ .
}

It follows from the linear system \eqref{hqq} that
\EQ{
\partial_\pm \newf \newf^{-1}=(1\pm\lambda)\partial_\pm\chi \chi^{-1}+  
\chi\Lambda_\pm\chi^{-1}\ .
\label{use2}
}
Since the left-hand side is independent of $\lambda$, the residues
of the right-hand side at $\lambda=\lambda_i$ and $\mu_i$ must
vanish,\footnote{In the following we will assume that
  $\lambda_i\neq\mu_j$ for any pair $i,j$. If the contrary is true
  then additional conditions must be imposed as we shall see later
  with an example.}
giving
\SP{
(1\pm\lambda_i)(\partial_\pm
Q_i)\Big(1+\frac{R_j}{\lambda_i-\mu_j}\Big)
+   Q_i\Lambda_\pm\Big(1+\frac{R_j}{\lambda_i-\mu_j}\Big)&=0\ ,\\
-(1\pm\mu _i)\Big(1+\frac{Q_j}{\mu_i-\lambda_j}\Big)\partial_\pm
R_i
+   \Big(1+\frac{Q_j}{\mu_i-\lambda_j}\Big)\Lambda_\pm R_i&=0\ ,
}
which are solved by 
\EQ{
(1\pm\lambda_i)\partial_\pm \BF_i^\dagger=-  
\BF_i^\dagger\Lambda_\pm\ ,\qquad
(1\pm\mu _i)\partial_\pm \BH_i= \Lambda_\pm \BH_i\ .
}
The solutions of these equations are
\EQ{
\BF_i=\big(\Psi_0(\lambda_i)^\dagger\big)^{-1}\Bvarpi_i\ ,\qquad
\BH_i=\Psi_0(\mu_i)\Bpi_i\ ,
}
for constant complex $n$-vectors $\Bvarpi_i$ and $\Bpi_i$. 

It can then be shown by tedious computation (re-produced in 
Appendix~\ref{AppSSSGEoM}) 
that the generic solution that arises from the
dressing procedure gives $\gamma$, as in \eqref{xxd} with $\Phi$ as in
\eqref{dph}, {\it i.e.\/}~
\EQ{
\gamma=\newf_0^{-1/2}\chi(+1)^{-1}\chi(-1)\newf_0^{1/2}
\label{cc1}
}
satisfies the equation-of-motion \eqref{eom}. Using the explicit
formulae for $\chi(\lambda)$ and its inverse, we find that the  
$G$-valued field of the reduction is 
\EQ{
\gamma=1-\frac2{(1-\mu_i)(1+\lambda_j)}\newf_0^{-1/2}\BH_i\big
(\Gamma^{-1}\big)_{ij}
\BF_j^\dagger \newf_0^{1/2}\ .
\label{oop}
}
So we see that the data of the dressing transformation constructs both
the sigma model magnon and the soliton in the SSSG.
There are simple formulae for the charges ${\cal Q}_{L,R}$ of the
dressed solution defined in
\eqref{cch}, and for ${\cal Q}_L$ the formula follows
directly from \eqref{use2}: since the right-hand side
is independent of $\lambda$ we can evaluate it at $\lambda=\infty$,
which gives
\EQ{
\partial_\pm{\cal F}{\cal F}^{-1}=\pm\partial_\pm \sum_iQ_i+
\Lambda_\pm\ .
\label{kaa1}
}
The final term here is precisely $\partial_\pm{\cal F}_0{\cal
  F}_0^{-1}$ and so the meaningful quantity to calculate is the charge
of the dressed solution relative  to the vacuum solution, and it follows
directly that
\EQ{
\Delta{\cal
  Q}_L=\sum_iQ_i\Big|_{x=\infty}-\sum_iQ_i\Big|_{x=-\infty}\
.
\label{use3}
}
In order to calculate the charge
${\cal Q}_R$, we use the fact that
\EQ{
{\cal F}^{-1}={\cal
  F}\Big|_{\lambda_i\to\lambda_i^{-1},\mu_i\to\mu_i^{-1},\Lambda_\pm\to
-\Lambda_\pm}\ ,
}
which can be proved directly. Hence, it follows that the charge
relative to the vacuum solution is
\EQ{
\Delta{\cal
  Q}_R=-\Delta{\cal Q}_L\Big|_{\lambda_i\to\lambda_i^{-1},
\mu_i\to\mu_i^{-1},\Lambda_\pm\to
-\Lambda_\pm}\ .
\label{use4}
}

Up till now we have described the B\"acklund transformation for
$SL(n,\C)$. However, as mentioned above we have to impose
involution conditions in order to describe a particular symmetric
space. As described in \cite{Harnad:1983we} for each choice of
symmetric space there are a set of involutions that must be
imposed. First of all, there is an involution 
(or possibly more than one) $\sigma_+$ that
picks out $F\subset SL(n,\C)$:
\EQ{
F=\big\{\newf\in SL(n,\C)\,\big|\,\sigma_+(\newf)=\newf\big\}\ .
}
Then there is a further involution
$\sigma_-$, that is detailed above for the explicit examples we have in mind, 
that picks out $F/G$ parameterized by ${\cal F}$ (as explained in
Appendix~\ref{Appscp}):
\EQ{
F/G\simeq\big\{\newf\in F\,\big|\,\sigma_-(\newf)=\newf^{-1}\big\}\ .
}
Notice also that the quotient group is identified as
\EQ{
G=\big\{\gamma \in F\,\big|\,\sigma_-(\gamma)=\gamma\big\}\ .
}
This allows us to prove that \eqref{cc1} is, as claimed, valued in
$G\subset F$. Using $\sigma_-(\newf_0^{\pm1/2})=\newf_0^{\mp1/2}$ and 
$\sigma_-(\chi(\pm1))=\newf^{-1}\chi(\pm1)\newf_0$ gives
\EQ{
\sigma_-(\gamma)=\newf_0^{1/2}\cdot \newf_0^{-1}\chi(+1)^{-1}\cdot
\newf^{-1}\chi(-1)\newf_0
\cdot \newf_0^{-1/2}=\gamma\ .
}

The involutions are each of of the following four types:
\SP{
&\sigma_1(\newf)=\theta \newf\theta^{-1}\ ,\qquad\sigma_2(\newf)=\theta 
\newf^*\theta^{-1}\ ,\\
&\sigma_3(\newf)=\theta(\newf^T)^{-1}\theta^{-1}\
,\qquad\sigma_4(\newf)=\theta{\newf^\dagger}^{-1}\theta^{-1}\ ,
\label{lsi}
}
where $\theta$ is either a symmetric, antisymmetric, hermitian or
anti-hermitian matrix. The involutions $(\sigma_1,\sigma_3)$ are
holomorphic while $(\sigma_2,\sigma_4)$ are anti-holomorphic. 

The correct way to impose these conditions on $\Psi(x;\lambda)$ 
are
\SP{
\Psi(\lambda)&=\sigma_+\big(\Psi(\tilde\lambda)\big)\ ,\\
\Psi(1/\lambda)&=\newf\sigma_-\big(\Psi(\tilde\lambda)\big)\ ,
\label{lo1}
}
where $\tilde\lambda=\lambda,\lambda^*$, if $\sigma_\pm$ is
holomorphic or anti-holomorphic, respectively.
Notice that if we take $\lambda=0$ and use 
$\newf=\Psi(0)$ and $\Psi(\infty)=1$, yields 
the correct conditions
\EQ{
\sigma_+(\newf)=\newf\ ,\qquad\sigma_-(\newf)=\newf^{-1}\ .
}
Furthermore it is easy to see that the vacuum solution \eqref{vso} satisfies
these conditions since $\Lambda_\pm\in{\mathfrak a}\subset{\mathfrak p}$.

Written in terms in terms of $\chi(\lambda)$ the conditions \eqref{lo1} become
\SP{
\chi(\lambda)&=\sigma_+\big(\chi(\tilde\lambda)\big)\ ,\\
\chi(1/\lambda)&=\newf\sigma_-\big(\chi(\tilde\lambda)\big)\newf_0^{-1}\ ,
}
which means that the two sets of poles $\{\lambda_i\}$ and $\{\mu_i\}$ must
be separately invariant under $\lambda\to \tilde\lambda$ (for $\sigma_+$), or
$\lambda\to 1/\tilde\lambda$ (for $\sigma_-$), 
for $\sigma_1$ and $\sigma_2$, and
mapping into each other for $\sigma_3$ and $\sigma_4$.

Rather than describe all the different cases, we specialize in this
work to the three examples (i)-(iii). 
Notice that in all our
examples the group $F$ is compact and the elements
$\Lambda_\pm^\dagger=-\Lambda_\pm$.
Since we are
principally interested in the application to string theory, there are
some additional conditions on $\Lambda_\pm$.  The vacuum 
solution ${\cal F}_0$ should be
a $t$-dependent, but $x$-independent, solution. This immediately
requires that $ \Lambda_+= \Lambda_-$. Notice that in cases
(ii) and (iii), the symmetric space has rank~1 and so either $\Lambda_+=\Lambda_-\equiv\Lambda$ or $\Lambda_+=-\Lambda_-\equiv\Lambda$, for a fixed element
$\Lambda$ (up to conjugation). For case (i), the principal chiral model, there
are more general models with $\Lambda_+\neq\Lambda_-$ that will be 
discussed elsewhere. 

The issue of relativistic invariance is quite subtle.
With the choice $\Lambda_+=\Lambda_-$, the vacuum solution is
$t$-dependent. Clearly, if we boost this solution then it will no
longer satisfy the Pohlmeyer constraints \eqref{laq}. This fact is
then inherited by the dressed solution. As we shall argue, although
the solution is localized in the sense, for example, 
that the density of its charges
${\cal J}^L_0$ and ${\cal J}^R_0$, {\it relative\/} to the vacuum, is
localized at a certain position in space moving with a certain
velocity, the solutions with different velocities are not related by
boosts. On the contrary, the solution in the reduced SSSG theory does
respect Lorentz transformations in the sense that the solutions with
different velocities are related by boosts (see Appendix~\ref{AppConserved}). 

Let us identify the velocity of the dressed solution.
The dependence on $x$ is via 
$\Psi_0(\xi)$, where $\xi$ is one of the $\lambda_i$ or $\mu_i$. The
localized nature of the soliton arises because when $\xi$ has a
imaginary part, $\Psi_0(\xi)$ has an exponential dependence on $x$.
Assuming that $\Lambda$ is anti-hermitian, the relevant
dependence is
\EQ{
\exp\Big[i\,\text{Im}\Big(\frac{x_+}{1+\xi}+\frac{x_-}{1-\xi}\Big)\Lambda
\Big]=\exp\Big[2i\,\text{Im}\Big(\frac{t-\xi x}{1-\xi^2}\Big)\Lambda\Big]
\label{hwp}
}
and this leads to 
exponential fall-off of the energy/charge density away from the centre
which is located at the solution of 
\EQ{
\text{Im}\Big(\frac{t-\xi x}{1-\xi^2}\Big)=0\ .
}
The velocity of the soliton is therefore
\EQ{
v=\frac{\text{Im}\,(1-\xi^2)^{-1}}{\text{Im}\,\xi(1-\xi^2)^{-1}}
=\frac{2r}{1+r^2}\cos\frac p2\ ,
\label{vel1}
}
where $\xi=re^{ip/2}$. Roughly speaking, the dressed solution describes $N$ solitons (for $i,j=1,\ldots,N$) and
$\lambda_i$ is a parameter that determines the velocity of the
$i^\text{th}$ soliton via \eqref{vel1} with $\xi=\lambda_i$. However,
for the cases $S^n$ and $\CP^n$, the additional constraints mean that
the solution actually represents less than $N$ independent solitons.

On the other hand, in the reduced SSSG model, the complete
dependence on~$t$ and~$x$ is through the combination
\EQ{
\newf_0^{-1/2}\Psi_0(\xi)=\exp\Big[\Big(\frac{(1+\xi^2)t}{1-\xi^2}
-\frac{2\xi x}{1-\xi^2}\Big)\Lambda\Big]\ .
\label{hwp2}
}
In this case, the model does have relativistic invariance and the
expression above can be written
\EQ{
\exp\Big[\Big(-t'
\sin\alpha-ix'\cos\alpha\Big)\Lambda\Big]\ ,
} 
where $(t',x')$ are the boosted coordinates defined in \eqref{boost}
and the parameter $\alpha$ and the rapidity $\vartheta$ are determined
by $r$ and $p$ as in \eqref{defa}.
The angle $\alpha$ sets both the size and
the internal angular velocity of the soliton. In the rest frame of the soliton,
$p=\pi$ or, equivalently, $\xi=ri$.

\section{The Principal Chiral Models}
\label{PCM}

As described in Section~\ref{SSSM}, we can either think of these theories as
symmetric space sigma models on $\ms=G\times G/G$ (so as a theory on $G\times G$
with an involution $\sigma_-$ that exchanges the two $G$
factors) or more directly as a sigma model defined on the Lie group $G$ (and thus not
needing a $\sigma_-$ involution).
We shall follow the second option and, hence, we formulate the theory
in terms of a $G$-valued field $\newf$ defined as a subgroup of $SL(n,\C)$ 
by the involution(s) $\sigma_+$, and in this approach there is no involution
$\sigma_-$.

In this work we will only consider the choice $G=SU(n)$, where there is a single involution 
\EQ{
\sigma_+(\newf)={\newf^\dagger}^{-1}
}
that, in the classification of~\cite{Harnad:1983we},
is of type $\sigma_4$ with $\theta=\BI$. 
Invariance under $\sigma_+$ requires 
\EQ{
\Psi(\lambda)={\Psi(\lambda^*)^\dagger}^{-1}\ ,
}
which is satisfied by imposing
$\mu_i=\lambda_i^*$ that in turn implies that
\EQ{
\BH_i=\BF_i\ ,\qquad \BK_i=\BX_i
\label{iv11}
}
in~\eqref{ResiduesMatrices}. This means that $\Bpi_i=\Bvarpi_i$ for each~$i$ 
and, moreover, that
\EQ{
\BX_i=\BK_i=\BF_j\big(\Gamma^{-1}\big)_{ji}\
,\qquad\Gamma_{ij}=\frac{\BF_i^\dagger \BF_j}{\lambda_i-\lambda_j^*}\>,
}
and
\EQ{
\chi(\lambda)=1+\frac{\BF_i\big(\Gamma^{-1}\big)_{ij}
\BF_j^\dagger}{\lambda-\lambda_j}\ ,\qquad
\chi(\lambda)^{-1}=1-\frac{\BF_i\big(\Gamma^{-1}\big)_{ij}
\BF_j^\dagger}{\lambda-\lambda_i^*}\ .
}
Then, using~\eqref{zlz} and~\eqref{drs}, the $SU(n)$ principal chiral model magnon is 
\EQ{
\newf=\chi(0)\newf_0=\newf_0-\frac{\BF_i\big(\Gamma^{-1}\big)_{ij}
\BF_j^\dagger \newf_0}{\lambda_j}\ ,
\label{PCMmagnon}
}
while, according to~\eqref{oop}, its solitonic avatar in the associated SSSG theory reads
\EQ{
\gamma= 1- \frac{2}{(1-\lambda_i^*)(1+\lambda_j)} \newf_0^{-1/2} \BF_i \big(\Gamma^{-1}\big)_{ij} \BF_j^\dagger \newf_0^{1/2}\>,
\label{PCMsoliton}
}
where
\EQ{
\BF_i = \Psi_0(\lambda_i^*) \Bvarpi_i\>.
}
In general, we will have to multiply~\eqref{PCMmagnon} and~\eqref{PCMsoliton} by constant phase factors
in order to enforce $\det\,\newf=1$ and $\det\,\gamma=1$, respectively.

The rank of the symmetric space $\ms=G\times G/G$ coincides with the rank of the Lie group $G$. Therefore, unless $G=SU(2)$, it gives rise to different Pohlmeyer reductions whose interpretation in the context of string theory is still to be understood. They are specified by two elements $\Lambda_\pm$ of the Cartan subalgebra of ${\mathfrak g}$ which, in the defining representation, are anti-hermitian diagonal matrices. In this work we will only consider the reductions corresponding to
\EQ{
\Lambda_+=\Lambda_-=i\,\text{diag}\big(\zeta_a\big)=\Lambda
\label{EqualLambda}
}
so that ${\cal F}_0$ only depends on~$t$. Then,
the vacuum solution of the associated linear system is
\EQ{
\Psi_0(\lambda)=\text{diag}\left(e^{\Theta_a(\lambda)}\right)\ ,
}
where
\EQ{
\Theta_a(\lambda)=
\frac{i\zeta_ax_+}{1+\lambda}+\frac{i\zeta_ax_-}{1-\lambda}= i\zeta_a \Bigl(\frac{2t}{1-\lambda^2} - \frac{2\lambda x}{1-\lambda^2}\Bigr)\ .
}
Furthermore, we will restrict ourselves to the cases with $\zeta_a\not=\zeta_b$ for $a\not=b$ so that $H^{(+)}=H^{(-)}=U(1)^{n-1}$,  
which correspond to the so-called (parity symmetric) homogeneous sine-Gordon
models~\cite{HSG}. 
Then, without loss of generality, we can order the $\zeta_a$ according to~\footnote{For those who are familiar with the HSG theories, this is equivalent to taking $\Lambda_+=\Lambda_-$ inside the principal Weyl chamber with respect to the standard choice of the basis of simple roots.}
\EQ{
\zeta_1> \zeta_2> \cdots \zeta_n\>.
}
More general reductions with $\Lambda_+\neq\Lambda_-$ will be discussed elsewhere. 

One-soliton solutions are obtained by considering a single pole in $\chi(\lambda)$, and so
\EQ{
\chi(\lambda)=1+\frac{\xi-\xi^*}{\lambda-\xi}
\frac{\BF\BF^\dagger}{\BF^\dagger \BF}\ ,
}
where 
\EQ{
\BF=\Psi_0(\xi^*)\Bvarpi
}
for a complex $n$-vector $\Bvarpi$. In terms of components 
\EQ{
F_a=e^{\Theta_a(\xi^*)}\varpi_a\ .
}
The complex $n$-vector $\Bvarpi$ represents a set of collective
coordinates for the solitons. Since $\chi(\lambda)$ and, hence, the soliton solutions are explicitly invariant under 
complex re-scalings $\Bvarpi\to\lambda \Bvarpi$, with $\lambda\in{\boldsymbol
  C}$, these collective coordinates span a $\CP^{n-1}$. Notice that constant shifts of the solitons in
space and time act on the collective coordinates via
\EQ{
\Bvarpi\longrightarrow \exp\Big[\frac{\delta
  x_+}{1+\xi^*}\Lambda+\frac{\delta x_-}
{1-\xi^*}\Lambda\Big]\Bvarpi\ .
\label{com}
}
So some of the collective coordinates fix the position of the soliton
in space and determine the temporal origin. The interpretation of the remaining ``internal'' collective coordinates will emerge when we analyze the solutions in more detail.

First of all, we think of the $SU(n)$ principal chiral model magnons.
Using~\eqref{PCMmagnon}, the group-valued field $\newf$ is given by
\EQ{
\newf_{ab}=e^{ip/n}\biggl(\delta_{ab}e^{\Theta_a(0)}-\frac{\xi-\xi^*}{\xi}
\frac{e^{\Theta_a(\xi^*)}\varpi_a\varpi_b^*
e^{-\Theta_b(\xi)+\Theta_b(0)}}
{\sum_c |\varpi_c|^2e^{\Theta_c(\xi^*)-\Theta_c(\xi)}}\biggr)\ ,
\label{PCMmagnonB}
}
where we have
multiplied $\newf$ by the phase $e^{ip/n}$ in order to enforce $\det\,\newf=1$.
This magnon carries $SU(n)_L\times SU(n)_R$ charges $\Q_{R/L}$ whose value relative to the vacuum solution can be calculated using~\eqref{use3} and~\eqref{use4}. The result is $\Delta\Q_{R/L}={\rm diag} \bigl(\Delta\Q_{R/L}^a \bigr)$ with
\SP{
&
\Delta\Q_L^a = -2i\> \big|r\sin\frac{p}{2}\big|\>\bigl(\delta_{a,{\rm min}} - \delta_{a,{\rm max}}\bigr)\\[5pt]
&
\Delta\Q_R^a = -2i\> \big|r^{-1}\sin\frac{p}{2}\big|\>\bigl(\delta_{a,{\rm min}} - \delta_{a,{\rm max}}\bigr)\>,
\label{PCMmagnonCharge}
}
where $\xi= re^{ip/2}$.
Next, we look at the solitonic avatar of the magnon~\eqref{PCMmagnonB} in the SSSG model which is provided by~\eqref{PCMsoliton}. It reads
\EQ{
\gamma_{ab}=e^{iC}\bigg(\delta_{ab}-\frac{2(\xi-\xi^*)}{(1-\xi^*)(1+\xi)}\cdot
\frac{e^{\Theta_a(\xi^*)-\Theta_a(0)/2}\varpi_a\varpi_b^*
e^{-\Theta_b(\xi)+\Theta_b(0)/2}}
{\sum_c |\varpi_c|^2e^{\Theta_c(\xi^*)-\Theta_c(\xi)}}\biggr)\ ,
\label{nnn1}
}
where we have multiplied $\gamma$ by the constant phase
\EQ{
e^{iC}= \Bigl(\frac{1-r^2+2ri\sin\frac{p}{2}}{1-r^2-2ri\sin\frac{p}{2}}\Bigr)^{1/n}
}
to enforce $\det\,\gamma=1$.
This field configuration satisfies the equations-of-motion \eqref{www1} with
$A^{(L)}_-=A^{(R)}_+=0$. Then, the value of the unambiguously well-defined Lorentz invariant SSSG charge $Q_L-Q_R$ carried by this soliton is provided by
\EQ{
\gamma(+\infty) \gamma^{-1}(-\infty) = e^{Q_L-Q_R}\>,
}
which follows from \eqref{GaugeSliceSSSG},~\eqref{LocalCharge}, and $\gamma_0^{vac}=1$.
The result is $Q_L-Q_R={\rm diag} \bigl(Q_L^a-Q_R^a \bigr)$ with
\EQ{
Q_L^a-Q_R^a = 2i \arctan\Bigl(\frac{2|r|}{r^2-1}\Bigr)\>\bigl(\delta_{a,{\rm min}} - \delta_{a,{\rm max}}\bigr)\>.
\label{PCMavatarCharge}
}
Finally, the mass of this SSSG soliton can be calculated using~\eqref{ener2}, which leads to
\EQ{
M=\frac{4|r|}{(r^2+1)} \>\bigl(\zeta_{\rm min}-\zeta_{\rm max}\bigr)
\>.
\label{PCMmass}
}
Eqs.~\eqref{PCMavatarCharge} and~\eqref{PCMmass} show that all the non-trivial SSSG solutions are obtained with $r>0$, and that charge conjugation corresponds to $r\rightarrow 1/r$.
Moreover, eqs.~\eqref{PCMmagnonCharge},~\eqref{PCMavatarCharge}, and~\eqref{PCMmass} unravel the r\^ole of the collective coordinates $\Bvarpi$. 
This solution is actually a superposition of ``max''--``min'' basic solitons, all mutually at rest, which exhibits that, with the special choice $\Lambda_+=\Lambda_-$, there are no forces between them. The basic solitons are
associated to the pairs  $(a,a+1)$, with only $\varpi_a$ and $\varpi_{a+1}$ non-vanishing. Those with a knowledge of root systems will appreciate
that these basic solitons are naturally associated to the
simple roots of $SU(n)$ and a particular $SU(2)\subset SU(n)$.\footnote{The composite nature of these solutions was already noticed in~\cite{FernandezPousa:1997iu} in the context of the homogeneous sine-Gordon theories.}
Then, the r\^ole of the collective coordinates $\varpi_a$ is to fix the relative space-time positions of the basic solitons.

Let us analyze the $SU(2)$ case in more detail, since $SU(2)\simeq S^3$ and, in any case,
this describes the basic solitons of the $SU(n)$ theory. We take
\EQ{
\zeta_1=-\zeta_2=\frac{1}{2}\; \Rightarrow \;\Lambda=\frac{i}{2}\>\text{diag}\bigl(1,-1\bigr)\ .
}
Shifting $x_\pm$ as in \eqref{com}, and using the overall scaling
symmetry, allows us to fix without-loss-of-generality
$\Bvarpi=(1,1)$. As with more
general solutions, $\xi=re^{ip/2}$ determines the velocity of the
soliton as well as the angular velocity of the internal motion, the
former as in \eqref{vel1}. The solution has the explicit form
\EQ{
\newf=\MAT{e^{i t}\bigl(\cos\frac{p}{2} + i\sin\frac{p}{2} \tanh(x'\cos\alpha)\bigr)\;&\;
-i\sin\frac{p}{2} e^{-it'\sin\alpha\>}\text{sech}(x'\cos\alpha)\\[5pt]
-i\sin\frac{p}{2} e^{+it'\sin\alpha\>}\text{sech}(x'\cos\alpha)\;
&\; e^{-i t}\bigl(\cos\frac{p}{2} - i\sin\frac{p}{2} \tanh(x'\cos\alpha)\bigr)
}\>,
\label{MagnonSU2}
}
where $x'=x\cosh\vartheta-t\sinh\vartheta$ and $t'=t\cosh\vartheta-x\sinh\vartheta$ are the boosted coordinates. The parameter $\alpha$ and the rapidity $\vartheta$ are determined by the
two parameters $p$ and $r$ via~\eqref{defa}. Notice,
that the moving solution is {\it not\/}
the boost of the solution at rest because of the $e^{\pm i t}$ factors.
Using~\eqref{PCMmagnonCharge}, the $SU(2)_L\times SU(2)_R$ charges relative to the
vacuum carried by this magnon can be written as
\EQ{
\Delta{\cal Q}_L=-4|r\sin\sigma|\Lambda\ ,\qquad
\Delta{\cal Q}_R=-4|r^{-1}\sin\sigma|\Lambda\ .
\label{chr1}
}
It is not difficult to check that~\eqref{MagnonSU2} corresponds to Dorey's dyonic magnon~\eqref{dm4}.\footnote{The explicit relationship reads
\EQ{
{\bf X}= -\text{Re}(\newf_{11}) {\bf e}_1+  \text{Im}(\newf_{11}) {\bf e}_2
-  \text{Im}(\newf_{12}) {\bf \Omega}^{(1)} - \text{Re}(\newf_{12}){\bf \Omega}^{(2)}\>.
}
}

Next we turn to the soliton avatar of~\eqref{MagnonSU2}. In the rest frame ($\xi=ri$, or $p=\pi$), it is
\EQ{
\gamma=\frac1{r^2+1}\MAT{r^2-1-2ri\tanh(x\cos\alpha)
\;&\; ire^{-2i t\sin\alpha}\text{sech}(x\cos\alpha)\\[5pt]
2ire^{+i t\sin\alpha}\text{sech}(x\cos\alpha)
&r^2-1+2ri\tanh(x\cos\alpha)}\ ,
\label{SolitonSU2}
}
along with $A_+^{(L)}=A_-^{(R)}=0$. Using~\eqref{defa} with $p=\pi$, in this equation
\EQ{
\cos\alpha =\frac{2r}{1+r^2} \>,\qquad \sin\alpha=\frac{1-r^2}{1+r^2}\>.
}
Then, the charge and mass carried by this SSSG soliton can be written as
\EQ{
Q_L-Q_R= 4 \arctan\Bigl(\frac{2|r|}{r^2-1}\Bigr)\Lambda\>, \qquad
M=\frac{8|r|}{(r^2+1)}
=4\big|\sin\big(\frac{1}{2}\Tr[\Lambda(Q_L-Q_R)]\bigr)\bigr|\>.
\label{ChargesCSG}
}
Eq.~\eqref{SolitonSU2} provides the well known one soliton solutions of the complex sine-Gordon equation~\cite{CSGsol}. Notice that the $r=1$ soliton is static. It is the embedding of the usual sine-Gordon soliton in the reduced $SU(2)$ principal chiral model. For this configuration, the  charge $Q_L-Q_R$ is uniquely defined only modulo $4\pi\Lambda$, a feature that played an important r\^ole in the construction of the CSG scattering matrix proposed in~\cite{Dorey:1994mg}.

In~\cite{Miramontes:2004dr}, it was shown that this SSSG soliton saturates a Bogomol'nyi-type bound, which explains the explicit relationship between mass and charge shown in~\eqref{ChargesCSG}. If we choose axial gauging, then $Q_L-Q_R$ corresponds to a $U(1)$ Noether charge and these solutions provide two-dimensional examples of $Q$-balls, which has been recently exploited to investigate some aspects of the dynamics of that type of extended solutions in quantum field theories~\cite{Bowcock:2008dn}.

\section{Complex Projective Space}
\label{CPn}

In this case the target space of the sigma model is the symmetric
space $SU(n+1)/U(n)$. As we have described in Section
\ref{dress}, this is picked out 
from the universal construction in $SL(n,\C)$ by two
involutions; $\sigma_+(\newf)={\newf^{\dagger}}^{-1}$, 
along with $\sigma_-$ in \eqref{iv3}. Notice that $\sigma_-$ is of type
$\sigma_1$ in the list \eqref{lsi} and is consequently holomorphic.
The vacuum solution is defined in \eqref{yhh}.

Turning to the dressing transformation,
invariance under $\sigma_-$ requires that 
\EQ{
\Psi(1/\lambda)=\newf\theta\Psi(\lambda)\theta^{-1}
}
and this means that the poles $\{\lambda_i\}$ must come in 
pairs $(\lambda_i,\lambda_{i+1}=1/\lambda_i)$ and we can think of a
single soliton as being a pair of the basic solitons of the
$SU(n)$ principal chiral model. In addition, the fact that the poles
come in pairs, requires associated conditions for each pair $i=1,3,\ldots$:
\EQ{
\Bvarpi_{i+1}=\theta\Bvarpi_i\ ,
}
which in turn means that
\EQ{
\BF_{i+1}=\Psi_0(1/\lambda_i^*)\theta\Bvarpi_i=\newf_0\theta
\Psi_0(\lambda_i^*)\Bvarpi_i
=\newf_0\theta \BF_i\ .
\label{gcd}
}

Let us consider in more detail the one soliton solution obtained from
a single pair of poles $\{\xi,1/\xi\}$. 
The dressing factor is 
\EQ{
\chi(\lambda)=1+\frac{Q_1}{\lambda-\xi}+\frac{Q_2}{\lambda-1/\xi}\ ,\\
}
and the matrix $\Gamma_{ij}$ has components
\SP{
&\Gamma_{11}=\frac{\beta}{\xi-\xi^*}\ ,\qquad
\Gamma_{12}=\frac{\xi^*\gamma}{|\xi|^2-1}\ ,\\
&\Gamma_{21}=-\frac{\xi\gamma}{|\xi|^2-1}\ ,\qquad
\Gamma_{22}=-\frac{|\xi|^2\beta}{\xi-\xi^*}\ ,
}
where we have defined the two real numbers
\EQ{
\beta=\BF^\dagger\BF\ ,\qquad\gamma=\BF^\dagger \newf_0\theta\BF\ ,
\label{dbg}
}
where $\BF\equiv\BF_1$. Therefore
\SP{
Q_1&=\frac1\Delta\Big[-\frac{|\xi^2|\beta}{\xi-\xi^*}\BF\BF^\dagger+
\frac{\xi\gamma}
{|\xi|^2-1}\newf_0\theta\BF\BF^\dagger\Big]\ ,\\
Q_2&=\frac1\Delta\Big[\frac{\beta}{\xi-\xi^*}\newf_0\theta
\BF\BF^\dagger\theta \newf_0^\dagger
-\frac{\xi^*\gamma}
{|\xi|^2-1}\BF\BF^\dagger\theta \newf_0^\dagger\Big]\ .
}
In the above, we have defined
\EQ{
\Delta=\det\,\Gamma=\frac{|\xi|^2\gamma^2}{(|\xi|^2-1)^2}
-\frac{|\xi|^2\beta^2}{(\xi-\xi^*)^2}\ .
}
The solution depends on the complex vector $\Bvarpi\equiv \Bvarpi_1$ and the
complex number $\xi$. In addition,
\EQ{
\chi(\lambda)^{-1}=1+\frac{R_1}{\lambda-\xi^*}+\frac{R_2}{\lambda-1/\xi^*}\
,
}
where
\SP{
R_1&=\frac1\Delta\Big[\frac{|\xi^2|\beta}{\xi-\xi^*}FF^\dagger+
\frac{\xi^*\gamma}
{|\xi|^2-1}FF^\dagger \theta \newf_0\Big]\ ,\\
R_2&=\frac1\Delta\Big[-\frac{\beta}{\xi-\xi^*}\newf_0\theta FF^\dagger
\theta \newf_0^\dagger
-\frac{\xi\gamma}
{|\xi|^2-1}\newf_0\theta FF^\dagger\Big]\ .
}

The magnon solution is obtained from
$\newf=\chi(0)\newf_0$. It corresponds to the projective
coordinates\footnote{This is similar to the Euclidean space formulae
  in \cite{Sasaki:1984tp}.}
\EQ{
\BZ=\big(\tilde\alpha+\theta\BF\BF^\dagger \theta\big)\BZ_0\ ,
}
where
\EQ{
\tilde\alpha=-\frac{\xi\beta}{\xi-\xi^*}-\frac\gamma{|\xi|^2-1}\ .
}

The complex $n+1$-vector $\Bvarpi$ represents a set of collective
coordinates for the magnon. In fact, it is easy to see that only this
vector up to complex re-scalings $\Bvarpi\to\lambda\Bvarpi$ lead to
inequivalent solutions. 
By making shifts in $x_\pm$, as in \eqref{com}, we can set always set, say,
$\varpi_2=0$ and then use the scale symmetry to set
$\varpi_1=i$,\footnote{The fact that we choose $i$ here will make it
  simpler to relate the solution to the case $\ms=S^n$.} so that
\EQ{
\Bvarpi=i\ee_1+\BOmega\ ,\qquad
\BOmega\cdot\ee_1=\BOmega\cdot\ee_2=0\>,
\label{ccc1}
}
where the constant vector 
$\BOmega$ is the internal collective coordinates of the 
magnon. The explicit solution is rather cumbersome to write down,
\EQ{
\BZ=Z_1\ee_1+Z_2\ee_2+Z_3\BOmega\ ,
\label{tfu}
}
where
\SP{
Z_1&=\tilde\alpha\cos t+\cos\left(\frac{-2e^{ip/2}rx+2e^{ip}t}
{e^{ip}-r^2}\right)\cos\left(\frac{-2e^{ip/2}rx+(r^2+1)e^{ip}t}
{-1+e^{ip}r^2}\right)\\
Z_2&=-\tilde\alpha\sin t-
\sin\left(\frac{-2e^{ip/2}rx+2e^{ip}t}
{e^{ip}-r^2}\right)\cos\left(\frac{-2e^{ip/2}rx+(r^2+1)e^{ip}t}
{-1+e^{ip}r^2}\right)\\
Z_3&=-
\cos\left(\frac{-2e^{ip/2}rx+(r^2+1)e^{ip}t}
{-1+e^{ip}r^2}\right)\ ,
}
and
\SP{
\tilde\alpha=&\frac{e^{ip}}{1-e^{ip}}\left[
|\BOmega|^2+\cos\left(\frac{4ir\sin\tfrac p2(-(r^2+1)x+2rt\cos\tfrac p2)}
{2r^2\cos p-1-r^4}\right)\right]\\
&+\frac1{1-r^2}\left[|\BOmega|^2-\cos\left(\frac{2(r^2-1)(2rx\cos\tfrac
    p2-(1+r^2)t)}{2r^2\cos p-1-r^4}\right)\right]
}
The magnon carries $SU(n)$ charge which can be extracted from
\eqref{use3}. The computation is simplified by noticing that
the off-diagonal elements in $Q_i=\BF_i(\Gamma^{-1})_{ij}\BF_j^\dagger$
(those with $j\neq i$) vanish as $x\to\pm\infty$ and so do not
contribute to the charge. This is because as $x\to\pm\infty$, $\beta$,
as defined in \eqref{dbg}, diverges exponentially, while $\gamma$,
also defined in \eqref{dbg}, remains bounded. The remaining
two contribution to the charge are then easily evaluated to give
\EQ{
\Delta {\cal Q}_L=-2\frac{1+r^2}r|\sin\frac p2\,|\Lambda\ .
}

The magnon solution that we have constructed above is
apparently singular when $|\xi|=1$, {\it i.e.}~$r=1$ or $\alpha=0$. 
However, a regular
solution in this limit can be constructed by imposing the additional
condition that
\EQ{
\gamma=F^\dagger \newf_0\theta F=\Bvarpi^\dagger \theta\Bvarpi=0\ ,
}
which can be written as a condition on the internal collective
coordinates,\footnote{In addition, it is necessary that $\varpi_2=0$.}
\EQ{
|\BOmega|=1\ .
}
In this case, the matrix $\Gamma$ is diagonal and the dressing
transformation has the simpler form:
\EQ{
\chi(\lambda)=1+\frac{\xi-\xi^*}{\lambda-\xi}\frac{\BF\BF^\dagger}{\beta}-
\frac{\xi-\xi^*}{\lambda-\xi^*}\frac{\newf_0\theta\BF\BF^\dagger
  \theta \newf_0^\dagger}{\beta}\ ,
}
The solution can also be obtained from \eqref{tfu} by setting
$|\BOmega|=1$ and taking the limit $r\to1$. It is not difficult to see
that up to a re-scaling by
\EQ{
-\frac{\cosh^2x'}{\sin\tfrac p2}\ ,
}
the solution is precisely an embedding of the Hofman-Maldacena
magnon in \eqref{hmm}. With reference to the discussion in Section \ref{GM},
it is the one associated to $\RP^2\subset\CP^n$.

The solitonic avatar of the magnon \eqref{tfu} 
in the SSSG theory is 
\footnote{In general 
this solution 
has $\det\,\gamma=e^{iC}$, for a
  constant $C$ and so in order that $\gamma\in G$ we should re-scale it by
  an appropriate compensating factor, as is done below in the explicit
  expressions.} 
\SP{
\gamma=1+\frac2\Delta &\Big[
\frac{|\xi^2|\beta}{(\xi-\xi^*)(1-\xi^*)(1+\xi)}
\newf_0^{-1/2}\BF\BF^\dagger \newf_0^{1/2}\\ &+
\frac{\xi\gamma}{(|\xi|^2-1)(1-1/\xi^*)(1+\xi)}\newf_0^{1/2}\theta\BF
\BF^\dagger
\newf_0^{1/2}\\ &+\frac{\xi^*\gamma}{(|\xi|^2-1)(1-\xi^*)(1+1/\xi)}
\newf_0^{-1/2}\BF\BF^\dagger\theta \newf_0^{-1/2}\\ &
-\frac{\beta}{(\xi-\xi^*)(1-1/\xi^*)(1+1/\xi)}
\newf_0^{1/2}\theta\BF\BF^\dagger\theta \newf_0^{-1/2}\Big]\ .
}
In the rest frame, $p=\pi$, this solution has the explicit form
\EQ{
\gamma=\left(\begin{array}{cc|c}\gamma_{11} & 0 & \B0^T\\ 
0 &\gamma_{22} & \gamma_{23}\,\BOmega^\dagger \\ \hline
\B0 & \gamma_{32}\,\BOmega &{\bf 1}+
(\gamma_{33}-1)\BOmega\BOmega^\dagger\end{array}\right)\ ,
\label{sal3}
}
where
\SP{
\gamma_{11}&=e^{2i\eta}
\frac{(r-i)^2|\BOmega|^2+2ir\cos2T+(r^2-1)\cosh2X}
{(r+i)^2|\BOmega|^2-2ir\cos2T+(r^2-1)\cosh2X}\ ,\\
\gamma_{22}&=e^{2i\eta}
\frac{(r-i)^2e^{-2i\eta}|\BOmega|^2-2ir\cos2T+(r^2-1)\cosh2X}
{(r+i)^2|\BOmega|^2+2ir\cos2T+(r^2-1)\cosh2X}\ ,\\
\gamma_{33}&=e^{-4i\eta/3}
\frac{(r+i)^2e^{2i\eta}|\BOmega|^2-2ir\cos2T+(r^2-1)\cosh2X}
{(r-i)^2|\BOmega|^2+2ir\cos2T+(r^2-1)\cosh2X}\ ,\\
\gamma_{23}&=
-e^{-i\eta/3}\frac{8r\sin(T+iX)}{(r-i)^2|\BOmega|^2+2ir\cos2T
+(r^2-1)\cosh2X}\ ,\\
\gamma_{32}&=
-e^{-i\eta/3}\frac{8r\sin(T-iX)}{(r-i)^2|\BOmega|^2+2ir\cos2T
+(r^2-1)\cosh2X}\ ,
}
where $e^{i\eta}=(r+i)/(r-i)$, and where
\EQ{
T=\frac{r^2-1}{r^2+1} t\ ,\qquad
X=\frac{2r}{r^2+1} x\ .
}
These solutions have vanishing SSSG charges $Q_L=Q_R=0$.
The mass of the solution can be computed using the
expression for the energy in \eqref{ener2} and one finds
\EQ{
M=\frac{8 r}{1+r^2}=4\cos\alpha\ .
}
Notice that it is more meaningful to write the result 
in terms of the parameter $\alpha$ defined in \eqref{defa}. The energy of
the general moving solution \eqref{ener2} is
\EQ{
{\mathscr E}=\frac{8 r}{1+r^2}\big|\sin\frac p2\big|\ .
}

The solution with $|\xi|=1$, is obtained by first
taking the limit $|\BOmega|\to1$ and then $r\to1$ (note these limits
do not commute). In this limit, and in the soliton rest frame,
\EQ{
\gamma=\left(\begin{array}{cc|c}-1 & 0 & \B0^T\\ 
0 &-1+2\,\text{sech}^2( x) & 
2\tanh( x)\,\text{sech}\,( x)\,\BOmega^\dagger \\ \hline
\B0 & 2\tanh( x)\,\text{sech}\,( x)\,\BOmega &
{\bf1}-2\,\text{sech}^2( x)\,\BOmega\BOmega^\dagger\end{array}\right)\ ,
\label{vxx1}
}
which is a static solution.

\section{The Spheres}
\label{Sn}

In this case the target space of the sigma model is the symmetric
space $S^n\simeq SO(n+1)/SO(n)$, and
the symmetric space is
picked out by the three involutions
\EQ{
\sigma_+^{(1)}(\newf)={\newf^{\dagger}}^{-1}\ ,
\qquad \sigma_+^{(2)}(\newf)=\newf^*\ ,\qquad
\sigma_-(\newf)=\theta \newf\theta^{-1}\ ,
}
where $\theta$ is given in \eqref{xx2}. Notice that $\sigma_-$ is of type
$\sigma_1$ in the list \eqref{lsi} and is consequently holomorphic.
The Pohlmeyer reduction is defined by taking $\Lambda_\pm$ as in
\eqref{cpp2}.
If we compare with the discussion of $\CP^n$ the only difference
is the reality condition $\newf^*=\newf$.

The simplest magnon solution is obtained by considering the dressing
transformation with a pair of poles $\xi$
and $1/\xi$, where $\xi$ is a phase. The constraints on the collective
coordinates are (with $\Bvarpi_1=\Bvarpi$)
\EQ{
\Bvarpi_2=\theta\Bvarpi\ ,\qquad\Bvarpi^*=\theta\Bvarpi\ ,\qquad 
\Bvarpi^\dagger
\theta\Bvarpi=0\ .
\label{ccc1b}
}
These are precisely the same conditions on the magnon of the $\CP^n$
case with $r=1$, with an additional reality condition. So just as in 
\eqref{ccc1} we have 
\EQ{
\Bvarpi=i\ee_1+\BOmega\ ,
\label{ccc2}
}
where now $\BOmega$ is a real unit vector orthogonal to $\ee_1$ and
$\ee_2$. Hence, the magnon has an internal collective coordinate
taking values in $S^{n-2}$. This magnon is precisely the
Hofman-Maldacena magnon \eqref{hmm}.
The soliton in the associated SSSG theory is precisely the $r=1$
solution in the $\CP^n$ case \eqref{vxx1} with the additional restriction
that $\BOmega$ is real.

\section{Dyonic Magnons/Solitons}
\label{Beyond}

One characteristic feature of the magnon/soliton solutions that we have
generated using the dressing transformation acting on the vacuum
solution is that they carry a non-trivial moduli space of 
internal collective coordinates
\cite{Hofman:2006xt,Spradlin:2006wk}. 

Usually when solitons have internal collective coordinates one expects
there are more general solutions 
for which the collective coordinates become time
dependent. For a static soliton, the resulting motion is 
simply geodesic motion on the moduli
space corresponding to a metric which is constructed from the inner
products of the associated zero modes. When the moduli space arises
from the action of a global symmetry then the metric will be invariant
under the symmetry.
The situation is familiar for BPS monopoles in
gauge theories. In this case the monopoles carry an internal $S^1$
moduli space which can be thought of as the $U(1)$ charge orientation of the 
monopole. A more
general solution, the dyon, exists where the angle parameterizing the $S^1$
rotates with constant angular velocity. An important lesson for our present
situation is that the dyon solution now carries electric charge as a
consequence of the motion. Finding the dyon is not easy because the
motion of the collective coordinate has a non-trivial back-reaction on
the original solution.

In the present context, it is important to understand the
action of the symmetries on the collective coordinates. First of all, recall
that the sigma model with target space a symmetric space has a global $F$ symmetry under which ${\cal F}\to
U{\cal F}\sigma_-(U^{-1})$, $U\in F$. 
Once the Pohlmeyer reduction is performed,
this symmetry corresponds to $f_\pm \to U f_\pm$, which leaves the
SSSG field $\gamma=f_-^{-1} f_+$ invariant. Notice that the vacuum
solution is invariant under the subgroup $H\subset G\subset F$.\footnote{We are
  assuming here that $H^{(\pm)}$, the subgroups of $G$ that commute with
  $\Lambda_\pm$, are equal to $H$ since in this paper we have
  $\Lambda_+=\Lambda_-$.} Hence, the transformations $U\in H$ on a
magnon have
a well defined action on 
the collective coordinates $\Bvarpi\to U\Bvarpi$, {\it i.e.\/}~$\BOmega\to
U\BOmega$ in the $\CP^n$ and $S^n$ cases.
On the other hand, the SSSG theory exhibits a global 
$H_L\times H_R$ symmetry that acts as $f_\pm\to
f_\pm h_\pm^{-1}$ or, equivalently, $\gamma\to h_-\gamma h_+^{-1}$,
where $h_\pm\in H$. In particular, the vector subgroup $\gamma\to
U\gamma U^{-1}$ of transformations leaves the vacuum invariant and
acts as a transformation on the soliton's collective coordinates in
the same way as above: $\Bvarpi\to U\Bvarpi$. So the symmetry group $H$
action on the collective coordinates can be interpreted in terms of a
transformation of both the magnon's and soliton's collective
coordinates where $H\subset F$ and $H\subset G$, respectively. This
symmetry will play an important r\^ole in fixing the geometry on the
moduli space of collective coordinates.

For example, for the cases
$\ms=\CP^n$, the general magnon/soliton, \eqref{tfu} and \eqref{sal3},
has an internal collective coordinate $\BOmega$ which is a complex
$n-1$ vector (presented as a $n+1$-vector orthogonal to $\ee_1$ and $\ee_2$). 
For the 
particular solution with $r=1$, we have the additional constraint 
$|\BOmega|=1$, so that the moduli space of collective coordinates is
$S^{2n-3}$. In both cases there is a natural action of $H=U(n-1)$ on
the moduli space. However, in this case the symmetry is not large enough to
completely fix the metric on the moduli space.

For the case with $\ms=S^n$ the soliton has a moduli space of
collective coordinates equal to $S^{n-2}$ parameterized by the real
unit length $n-1$ vector 
$\BOmega$ (again presented as an $n+1$-vector orthogonal
to $\ee_1$ and $\ee_2$) on which there is a natural
action of $H=SO(n-1)$. In this case, the symmetry fixes the metric on
the moduli space (up to overall scaling).
Dorey's solution is precisely the dyon associated to the
Hofman-Maldacena magnon for the case $\ms=S^3=SO(4)/SO(3)$. 
In this case $\BOmega$
is a unit 2-vector in the subspace spanned by $\ee_3$ and
$\ee_4$. Allowing it to rotate with constant angular velocity,
\EQ{
\BOmega(t)=\cos(t\sin\alpha)\ee_3+\sin(t\sin\alpha)\ee_4\ ,
\label{baa}
}
leads to Dorey's dyonic magnon. However, in order to compute the
complete back-reacted solution, it is more convenient to notice that 
there is another realization of the $S^3=SO(4)/SO(3)$ model 
as the principal chiral model for $G=SU(2)$. The explicit map is
\EQ{
{\cal F}=\MAT{X_1+iX_2&iX_3+X_4 \\ iX_3-X_4&X_1-iX_2}\in SU(2)\ .
}
In the $SU(2)$
formulation, the dyonic magnon is just the {\it ordinary\/} magnon
solution which we described in Section \ref{PCM}. In particular, we wrote
\eqref{baa} in such a way that the parameter $\alpha$ is the same as
the one that appears as a parameter of the $SU(2)$ magnon. 

The dyonic magnon gives a dyonic generalization
of the SSSG soliton as we described in Section \ref{GM} for the more
general case with $\ms=S^n$, $n>3$. In particular, the
solution has non-trivial gauge fields $A_L^{(+)}$ and $A_R^{(-)}$, and, 
as we also explained in Section \ref{GM}, the dyonic solution can also be
embedded in $\CP^n$, for $n\geq3$ by using the maps
$S^3\to\RP^3\to\CP^n$. 
However, because the symmetry $H=U(n-1)$ is not large enough to fix
the metric on $S^{2n-3}$ there should exist another 
inequivalent class of dyonic solutions.
In more detail, invariance under $U(n-1)$ fixes the metric to be a
  linear combination 
\EQ{
ds^2=d\BOmega^\dagger\cdot d\BOmega+\xi(d\BOmega^\dagger\cdot
\BOmega-\BOmega^\dagger\cdot d\BOmega)^2\ ,
\label{met2}
}
up to overall scaling. When $\xi=0$ we have the usual spherically
symmetric metric and in this case there are no new dyon
solutions. However, when $\xi\neq0$, the new class of dyon solutions
are associated to geodesics of the form
\EQ{
\BOmega(t)=e^{h t}\Bp\ ,
}
where we can choose the overall orientation so that
$\Bp=(1,0,\ldots,0)$. The allowed algebra element $h$ 
can be found by solving the
geodesic equations for the metric \eqref{met2}. There are two classes
of solution, firstly 
\EQ{
h=i\left(\begin{array}{cc} 0& \Bw^\dagger \\ \Bw &
    0\end{array}\right)\ ,
}
where $\Bw$ is a complex $n-2$ vector. 
These give the embeddings of
Dorey's dyon. The new class corresponds to
\EQ{
h=i\left(\begin{array}{cc} 
v\;\;&\; \Bw^T \\[5pt]
 \Bw \;\;&\; (8\xi-1)v\Bw\,\Bw^T/|\Bw|^2\end{array}\right)\ ,
}
where $v$ is a real number and $\Bw$ is a real $n-2$ vector. This new
class includes the simple example
\EQ{
\BOmega(t)=\big(e^{ivt},0,\ldots,0\big)\ .
\label{nge2}
}
Such dyons will carry charge lying in the abelian subalgebra defined
by $h$. For example in the case $\ms=\CP^2$ considered at the end of
Section \ref{Pohl}, $\BOmega=\Omega$ is just a complex 1-vector (or number)
and only the new class of dyons with $\Omega(t)=e^{ivt}$  
will exist. In terms of the Lagrangian formulation via axial gauging
in \eqref{axiallag}, the dyon will correspond to a solution
for which $\tilde\psi=vt$. Finding the back-reaction on the fields
$\varphi(x)$ and $\theta(x)$ is a difficult challenge that we will not
solve here.

\section{Conclusions and outlook}
\label{Conclusions}

In this work we have considered the interplay between the magnons in the
sigma model describing string motion of certain symmetric spaces and
the solitons of the related SSSG equations. A notable result is that
the dressing procedure produces the magnon and soliton at the same
time without the need to implement the complicated map between the
two systems. We have also described how the dressing procedure in its
current understanding cannot
produce the more general dyonic magnon/soliton solutions which 
involve the non-trivial motion of the internal collective coordinates.
It would be interesting to try to find a generalization of the
dressing method which produces such dyonic solutions directly from the
vacuum. 
In this work we have restricted ourselves to the simplest compact
symmetric spaces and also to the simplest single magnon/soliton
solutions: generalizations will be presented elsewhere.

\acknowledgments

\noindent
JLM thanks the Galileo Galilei Institute for Theoretical Physics for  
the hospitality and the INFN for partial support while this work was in progress.
His work was partially  supported by MICINN (Spain) 
and FEDER (FPA2008-01838 and 
FPA2008-01177), by Xunta de Galicia (Consejer\'\i a de 
Educaci\'on and PGIDIT06PXIB296182PR), and by the 
Spanish Consolider-Ingenio 2010
Programme CPAN (CSD2007-00042).

\noindent TJH would like to acknowledge the support of STFC grant
ST/G000506/1 and the hospitality of the Department of 
Particle Physics and IGFAE at the University of Santiago de Compostela.

\startappendix

\Appendix{Relation to the Gauged Sigma Model Approach}
\label{AppSigmaModel}

In this appendix we summarize the relationship between the approach
described in~\cite{Miramontes:2008wt} (see
also~\cite{Eichenherr:1981sk}) and the formulation of the $F/G$
symmetric space sigma model used in Section~\ref{SSSM} in terms of the
principal chiral model for~$F$. 
In~\cite{Miramontes:2008wt}, the $F/G$ symmetric space sigma model is
formulated with two fields $\oldf  \in F$ and $B_\mu\in{\mathfrak g}$
subject to the gauge symmetry 
\EQ{
\oldf \rightarrow \oldf  g^{-1}\>, \quad
B_\mu \rightarrow g(B_\mu +\partial_\mu)g^{-1}\>,\quad g\in G\>.
\label{GaugeTrans}
}
If the Lie group $F$ is simple, the nonlinear sigma model is defined by the Lagrangian
\begin{equation}
\LAG= -\frac{1}{2\kappa} \mathop{\rm Tr} \bigl(J_\mu J^\mu\bigr)\>,
\label{LagSM}
\end{equation}
where the current $J_\mu = \oldf ^{-1}\partial_\mu \oldf -B_\mu
\rightarrow g J_\mu g^{-1}$ is covariant under gauge transformations. 

The relationship between the two formulations relies on the fact that
the solution space of the $F/G$ sigma model can be realized as a
subspace of the solution space of the $F$ principal chiral model,
which is a consequence of the following result due to
Cartan~\cite{Eichenherr:1979hz}: The smooth mapping
\EQ{
\Phi: F/G\rightarrow F\>,\quad {\rm with}\quad
\oldf G \mapsto \Phi(\oldf G)= \sigma_-(\oldf )\oldf ^{-1}\>,
\label{Map}
}
is a local diffeomorphism of $F/G$ onto the closed  totally geodesic
submanifold $M=\{\oldf \in F: \sigma_-(\oldf )=\oldf ^{-1}\}$, where
$\sigma_-$ is the involution of $F$ that fixes $G\subset F$ and gives
rise to the canonical decomposition~\eqref{CanonicalDec}. Examples of this map can be found in Appendix~\ref{Appscp}. Taking~\eqref{Map}
into account, the explicit connection between the two models was
worked out in~\cite{Eichenherr:1979hz} making use of  the
gauge-invariant field 
\EQ{
\newf=\sigma_-(\oldf )\oldf ^{-1}
\label{NewField}
}
that trivially satisfies the constraint~\eqref{cdo}; namely,
$\sigma_-(\newf )=\newf ^{-1}$. 
Notice that in~\eqref{LagSM} the gauge fields $B_\mu$ are just
Lagrangian multipliers whose equations-of-motion are
$J_\mu\big|_{{\mathfrak g}} =0$, 
which is equivalent to
\EQ{
B_\mu=\oldf ^{-1}\partial_\mu \oldf \big|_{{\mathfrak g}}
\quad{\rm and}\quad
J_\mu=\oldf ^{-1}\partial_\mu \oldf \big|_{{\mathfrak p}}\>.
}
Then, it is easy to check that 
\EQ{
{\cal J}_\mu =\partial_\mu \newf  \newf ^{-1} = -2\sigma_-(\oldf ) J_\mu \sigma_-(\oldf ^{-1})\>,
\label{TwoCurrents}}
and the Lagrangian~\eqref{LagSM} becomes
\EQ{
\LAG= -\frac{1}{2\kappa} \mathop{\rm Tr}\bigl(J_\mu J^\mu\bigr)
=-{1\over8\kappa}  \mathop{\rm Tr}\bigl({\cal J}_\mu {\cal J}^\mu\bigr)\>,
}
which is the Lagrangian of the $F$ principal chiral model subject to the constraint~\eqref{cdo}. Moreover, using the identity 
\EQ{
D_\mu J_\nu = \partial_\mu J_\nu +[B_\mu,
J_\nu]=-\frac{1}{2}\sigma_-(\oldf ^{-1}) \left( \partial_\mu {\cal
    J}_\nu - {1\over2}[{\cal J}_\mu, {\cal J}_\nu] \right)
\sigma_-(\oldf )\>, 
}
the equations-of-motion of the $F/G$ symmetric space sigma model become
\SP{
&D_{\pm} J_{\mp}=0 \>\Rightarrow\>
\partial_{\pm} {\cal J}_{\mp}-{1\over2} \bigl[{\cal J}_\pm,{\cal
  J}_\mp\bigr]=0, 
}
which are just~\eqref{wee}.

Now, taking~\eqref{TwoCurrents} into account, the constraints that
specify the Pohlmeyer reduction of the model in terms of constrained
principal chiral model field $\newf$ can be imported directly
from the Eqs.~(3.11) and (3.17) of \cite{Miramontes:2008wt}: 
\EQ{
\partial_\pm \newf \newf^{-1} = -2\sigma_-(\oldf )\> J_\pm\>
\sigma_-(\oldf ^{-1}) 
=
-2 \sigma_-(\oldf )\Bigl(\overline{g}_\pm
\bigl(\mu_\pm\Lambda_\pm\bigr)\overline{g}_\pm^{-1}\Bigl)\sigma_-(\oldf
^{-1})\>, 
} 
where $\overline{g}_\pm\in G$, $\Lambda_\pm\in{\mathfrak a}$, and
${\mathfrak a}$ is a maximal abelian subspace of ${\mathfrak p}$
in~\eqref{CanonicalDec}. They correspond to~\eqref{laq} with 
$f_\pm=\sigma_-(\oldf ) \overline{g}_\pm\in F$ 
where, for simplicity
and without loss of generality, we have fixed
$\mu_\pm=-\frac{1}{2}$.
One can think of these overall scales multipliers as having been
absorbed into $\Lambda_\pm$.
Moreover, since $\overline{g}_\pm\in G$, it is
straightforward to check that $\sigma_-(f_\pm)= \newf^{-1} f_\pm$, and
that $\gamma=\overline{g}_-^{-1}  \overline{g}_+ =f_-^{-1} f_+$ takes
values in $G$, in agreement with~\eqref{jjj} and~\eqref{ValuedinG},
respectively. 

\Appendix{Integrability, Conserved Currents, Energy-Momentum Tensor and Lorentz Transformations}
\label{AppConserved}

In order to uncover the integrability of the SSSG equations
\eqref{www1}, it is useful to formulate them as the zero curvature
condition
\EQ{
[{\cal L}_+,{\cal L}_-]=0\ ,
\label{ZeroApp}
}
with the components of the Lax operator ${\cal L}_\mu$ given by
\SP{
{\cal L}_+&=\partial_+ + \gamma^{-1}\partial_+\gamma + \gamma^{-1}
A_+^{(L)} \gamma -\frac{1}{2} z  \Lambda_+ \equiv{\cal L}_+
(x_\pm,\gamma,A_+^{(L)};z)\ , \\
{\cal L}_-&=\partial_- + A_-^{(R)} -\frac{1}{2} z^{-1}  \gamma^{-1}
\Lambda_-\gamma\equiv{\cal L}_- (x_\pm,\gamma,A_-^{(R)};z)\ .
\label{LaxOp}
}
In the above $z$, the {\sl spectral parameter\/}, is an arbitrary
auxiliary parameter whose introduction plays a key
r\^ole in establishing the integrability of the theory.
The zero curvature condition gives 
rise to an infinite number of conserved densities labeled by
their spin. The ones corresponding to spin~1 and 2 provide the usual
Noether currents and the components of the stress-energy tensor,
respectively. It is important to recall that the zero curvature
condition is subject to the gauge symmetry transformations 
\EQ{
\gamma \rightarrow h_-\> \gamma\> h_+^{-1}, \quad A_-^{(R)}\rightarrow h_+\bigl( A_-^{(R)}  + \partial_-\bigr) h_+^{-1}, \quad A_+^{(L)}\rightarrow h_- \bigl(A_+^{(L)}  + \partial_+\bigr) h_-^{-1}\>.
\label{ESSSG1}
}

We can deduce the form of those conserved densities using the
``Drinfeld-Sokolov procedure''~\cite{DS}. 
In order to do that, we notice that, with the introduction of the
spectral parameter, the Lax operator can be written in terms of 
the affine algebra
\EQ{
{\mathfrak f}^{(1)} =\sum_{k\in \boldsymbol{Z}} \Bigl( z^{2k} \otimes
{\mathfrak g} + z^{2k+1} \otimes {\mathfrak p}\Bigr) =\bigoplus_{k\in
  \boldsymbol{Z}}\> {\mathfrak f}^{(1)}_k
}
by means of
\EQ{
z\Lambda_+\equiv z\otimes \Lambda_+\in {\mathfrak f}^{(1)}_1, \quad
z^{-1}\Lambda_- \equiv z^{-1}\otimes \Lambda_-\in {\mathfrak f}^{(1)}_{-1},
\quad
A_\pm^{(L/R)}\equiv 1\otimes A_\pm^{(L/R)}\in {\mathfrak f}^{(1)}_{0}.
}
Moreover, $\gamma$ takes values in $G$ that is the group associated to the Lie algebra ${\mathfrak f}^{(1)}_{0}$.
Next,  we introduce $\Phi_{(+)}\in \exp ({\mathfrak f}^{(1)}_{<0})$, and solve
\EQ{
\Phi_{(+)} \Bigl(\partial_+ + \gamma^{-1}\partial_+\gamma + \gamma^{-1} A_+^{(L)} \gamma - \frac{1}{2} z\Lambda_+\Bigr)\Phi_{(+)}^{-1} = \partial_+  - \frac{1}{2} z\Lambda_+ + h^{(+)}, 
\label{DS}}
with
\EQ{
h^{(+)}=\sum_{k\leq0} z^{-k} h^{(+)}_{-k}\in {\rm Ker}\bigl({\rm Ad} (\Lambda_+)\bigr) \cup  {\mathfrak f}^{(1)}_{\leq0}.
}
Correspondingly,
\EQ{
\Phi_{(+)} \Bigl(\partial_- + A_-^{(R)} - \frac{1}{2} z^{-1}\gamma^{-1} \Lambda_-\gamma\Bigr)\Phi_{(+)}^{-1} = \partial_-  + I^{(+)}\>, \qquad I^{(+)} \in {\mathfrak f}^{(1)}_{\leq0}\>.
\label{DS2}
}
Then, the zero curvature condition implies
\EQ{
\bigl[\partial_+  - \frac{1}{2} z\Lambda_+ + h^{(+)},  \partial_-  + I^{(+)}\bigr]=0\>,
\label{Zero2}
}
The components of $h^{(+)}$ and $I^{(+)}$ on ${\rm Cent}\bigl({\rm Ker}\bigl({\rm Ad} (\Lambda_+)\bigr)$ provide an infinite set of local conserved densities, while the other components provide non-local conserved ones.
A second set of conserved quantities can be constructed starting from
\EQ{
\gamma\bigl(\partial_- + A_-^{(R)} -\frac{1}{2} 
z^{-1}\gamma^{-1} \Lambda_-\gamma\bigr)\gamma^{-1}= \partial_- -\partial_-\gamma\gamma^{-1}+ \gamma A_-^{(R)}\gamma^{-1} -\frac{1}{2}  z^{-1}\Lambda_-
\label{DSMinus}
}
instead of ${\cal L}_+$.

The explicit expression of the densities of spin~1 and~2 can be found by writing
\EQ{
\Phi_{(+)}=\exp\Bigl(\sum_{k\geq1}z^{-k}y_{-k} \Bigr), \qquad
z^{-k}y_{-k}\in {\mathfrak f}^{(1)}_{-k}\>,
}
and looking at the first components of~\eqref{DS}, which read
\AL{
&
h_0^{(+)} -\frac{1}{2}[\Lambda_+,y_{-1}]=\gamma^{-1}\partial_+\gamma + \gamma^{-1} A_+^{(L)} \gamma\equiv q \label{Recurr1}\\[5pt]
&
h_{-1}^{(+)}-\frac{1}{2}[\Lambda_+,y_{-2}] = -\partial_+ y_{-1} +[y_{-1},q]-\frac{1}{4}[y_{-1},[y_{-1},\Lambda_+]] \label{Recurr2}\\
&
\cdots\cdots\cdots\cdots\cdots\cdots\nonumber
}
Using~\eqref{Orthogonal}, eq.~\eqref{Recurr1} provides
\EQ{
y_{-1}\in {\rm Im}\bigl({\rm Ad} (\Lambda_+)\bigr)\>, \qquad h_0^{(+)}= {\bf P}_{{\mathfrak h}_+}\bigl(\gamma^{-1}\partial_+\gamma + \gamma^{-1} A_+^{(L)} \gamma\bigr)=A_+^{(R)}\>,
}
where we have also used~\eqref{dmi}.
In turn,~\eqref{DS2} gives
\EQ{
I^{(+)}_0= A_-^{(R)}.
 }
Therefore, the 0-grade component of~\eqref{Zero2} on ${\rm Ker}\bigl({\rm Ad} (\Lambda_+)\bigr)$ reads
\EQ{
\bigl[\>\partial_+ +A_+^{(R)},\> \partial_- + A_-^{(R)}  \>\bigl]
=0\ ,
}
which is one of the two equations in~\eqref{Flatness}. The other is obtained is a similar way starting from~\eqref{DSMinus} instead of ${\cal L}_+$. The local and non-local conserved quantities provided by these equations and their interpretation are extensively discussed in Section \ref{Pohl}.

The components of the stress-energy tensor are found by looking at the components of $h^{(+)}_{-1}$ and $I^{(+)}_{-1}$along $\Lambda_+$. Using~\eqref{Recurr2},
\EQ{
{\rm Tr}\bigl(\Lambda_+ h_{-1}^{(+)}\bigr)={\rm Tr}\Bigl(\Lambda_+\bigl([y_{-1},q]-\frac{1}{4}[y_{-1},[y_{-1},\Lambda_+]]\bigr)\Bigr)=-{\rm Tr}\Bigl( (q-h_0^{(+)})^2 \Bigr)\equiv 2T_{++}.
}
Correspondingly,~\eqref{DS2} provides
\EQ{
{\rm Tr}\bigl(\Lambda_+ I_{-1}^{(+)}\bigr)=-\frac{1}{2}{\rm Tr}\Bigl(\Lambda_+\gamma^{-1}\Lambda_- \gamma \Bigr)\equiv -2T_{-+}\>,
}
and~\eqref{Zero2} leads to
\EQ{
\partial_+T_{-+}+
\partial_-T_{++}=0.
}
The component $T_{--}$ is obtained is a similar fashion starting from~\eqref{DSMinus} instead of ${\cal L}_+$.
Then, the complete set of components of the energy-momentum tensor can be written as
\AL{
T_{++}&=-\frac{1}{2}{\rm Tr}\Bigl( (q-h_0^{(+)})^2 \Bigr)=-\frac{1}{2}\Tr\Big[\bigl(\partial_+\gamma\gamma^{-1}+ A_+^{(L)}\bigr)^2-{A_+^{(R)}}^2\Bigr]
\label{Tplusplus}
\\[5pt]
T_{--}&=-\frac{1}{2}\Tr\Big[\bigl(\gamma^{-1}\partial_-\gamma- A_-^{(R)}\bigr)^2 -{A_-^{(L)}}^2\Bigr]
\label{Tminusminus}
\\[5pt]
T_{-+}&=T_{-+}=+\frac{1}{4}{\rm Tr}\Big[\Lambda_+\gamma^{-1}\Lambda_-\gamma
\Big]\>,
\label{ener}
}
and it can be easily checked that these expressions are gauge invariant.

The formulation in terms of the Lax operator ${\cal L}_\pm$ is also useful to discuss the behaviour of the reduced equations under Lorentz transformations.
The SSSG equations \eqref{www1} 
are Lorentz invariant, which means that given a solution
\begin{equation}
\gamma=\gamma(x_+,x_-), \quad A_+^{(L)} =A_+^{(L)} (x_+,x_-), \quad A_-^{(R)}=A_-^{(R)}(x_+,x_-)
\end{equation}
we can generate a boosted one by simply
\SP{
&\gamma\rightarrow \gamma_\lambda=\gamma(\lambda^{-1} x_+,\lambda
x_-)\ ,\\ 
&A_+^{(L)}\rightarrow {A_+^{(L)}}_\lambda=\lambda^{-1} A_+^{(L)}
(\lambda^{-1} x_+,\lambda x_-)\ ,\\ &A_-^{(R)}\rightarrow
{A_-^{(R)}}_\lambda=\lambda^{+1} A_-^{(R)}(\lambda^{-1} x_+,\lambda
x_-)\ . 
}
This is equivalent to saying 
that the zero-curvature condition is invariant under the transformations
\begin{equation}
x_\pm\rightarrow \lambda^{\pm 1}x_\pm, \quad \gamma\rightarrow\gamma, \quad A_+^{(L)}\rightarrow \lambda^{-1}A_+^{(L)}
, \quad A_-^{(R)}\rightarrow \lambda^{+1} A_+^{(R)}\>.
\label{LorentzTransf}
\end{equation}
Correspondingly, the Lax operators \eqref{LaxOp}
transform as
\begin{eqnarray}
&&
{\cal L}_+ (x_\pm,\gamma,A_+^{(L)};z)\rightarrow
\lambda^{-1} {\cal L}_+ (x_\pm,\gamma,A_+^{(L)};\lambda z),\nonumber\\[5pt]
&&
{\cal L}_- (x_\pm,\gamma,A_-^{(R)};z)\rightarrow
\lambda^{+1}  {\cal L}_- (x_\pm,\gamma,A_-^{(R)};\lambda z).
\label{LorentzLax}
\end{eqnarray}
In other words, the Lorentz transformation~\eqref{LorentzTransf} is equivalent to the re-scaling of the spectral parameter $z\to\lambda z$, and the zero-curvature condition is invariant because it does not depend on~$z$.
Then, in~\eqref{DS} the Lorentz transformation~\eqref{LorentzLax}
induces the following transformation on the
conserved densities:
\begin{equation}
h^{(+)}_{-j}\rightarrow \lambda ^{-1-j} h^{(+)}_{-j}
\end{equation}
which, in particular, shows that $h^{(+)}_{0}$ 
is of spin~1 (currents) and, therefore, that the corresponding
conserved charges are Lorentz invariant.

In contrast to the SSSG equations, the Pohlmeyer reduced sigma model is not Lorentz invariant, as a consequence of the constraints~\eqref{laq}. However, 
we can use the formulation of the former in term of the Lax operators ${\cal L}_\pm$ to deduce a formal expression for the action of Lorentz transformations on the solutions to the reduced sigma model equations-of-motion. 
Consider the solutions to the $z$-dependent auxiliary linear problem
\begin{equation}
{\cal L}_+ (x_\pm,\gamma,A_+^{(L)};z) \Upsilon^{-1}(z) = {\cal L}_- (x_\pm,\gamma,A_-^{(R)};z^{-1})\Upsilon^{-1}(z)=0
\label{LinearProblem}
\end{equation}
where $\Upsilon(z)\equiv \Upsilon\bigl(x_\pm,\gamma,A_+^{(L)},A_-^{(R)};z \bigr)$, whose integrability conditions are provided by the zero-curvature
equation~\eqref{ZeroApp}.
As explained in~\cite{Miramontes:2008wt}, in the gauged sigma model approach the reduced sigma model configuration corresponding to a
given SSSG solution $\bigl\{ \gamma, A_+^{(L)}, A_-^{(R)}\bigr\}$ is specified by the solution to \eqref{LinearProblem} for $z=1$; namely, $f=\Upsilon(1)$.
Then,~\eqref{LorentzLax} shows that under a Lorentz transformation
$\Upsilon(z)\rightarrow \Upsilon(\lambda z)$, which induces the following transformation of the reduced sigma model configuration:
\EQ{
f =\Upsilon(1) \longrightarrow f_\lambda=\Upsilon (\lambda)\>.
}

\Appendix{The Spheres and Complex Projective Spaces}
\label{Appscp}

In this appendix, we explain how to map the spaces $S^n$ and
$\CP^n$, expressed in terms of their usual coordinates, into the group valued field ${\cal F}$ given by~\eqref{NewField}. 

A generic $f\in SO(n+1)$ satisfies
$f f^T= 1$ which is equivalent to $f_{ac} f_{bc}=\delta_{ab}$.
Then,
\EQ{
{\cal F} = \sigma_-(f) f^{-1} =\theta f\theta f^T 
}
which, in terms of components, reads
\EQ{
{\cal F}_{ab} =\theta_{ac} \Bigl(\delta_{cb} -2f_{c1} f_{b1} \Bigr).
\label{Fcal}
}
Now, for a symmetric space $\ms=F/G$, we have to use that $F=I_0(\ms)$
is the identity component of the group of isometries of $\ms$,
and that it acts transitively on $\ms=F/G$. 
This means that $\ms=F\cdot \Bp_0$ for an arbitrary point $\Bp_0\in
\ms$ and, moreover, that $G$ is the isotropy group (or little
group) of $\Bp_0$.
In our case, for $S^{n}=SO(n+1)/SO(n)$ we can take $\Bp_0=(1,0,\ldots,0)$, 
so that the point corresponding to $f$ is 
\EQ{
\BX = f\cdot\Bp_0 \;\Rightarrow\; X_a = f_{a1}
}
Then,~\eqref{Fcal} becomes
\EQ{
{\cal F} = \theta \Bigl(1 - 2 \BX\BX^T \Bigr),
\label{a11}
}
which is the parameterization we are looking for in terms of the unit
vector $\BX$, $|\BX|=1$. Notice that the map $\BX\to{\cal F}$, which provides a particular example of~\eqref{Map}, 
is surjective but not injective.

A similar argument can be followed for the case of the complex
projective spaces, in which case \eqref{a11} is replaced by 
\EQ{
{\cal F} = \theta \Bigl(1 - 2\frac{\BZ\BZ^\dagger}{|\BZ|^2} \Bigr)\ ,
\label{a12}
}
where $\BZ$ is a vector whose components are the 
usual $n+1$ projective coordinates of $\CP^n$. In this case, the map $\BZ\rightarrow \newf$ is one-to-one.

\Appendix{The SSSG Equations-of-Motion}
\label{AppSSSGEoM}

In this appendix we prove that the dressing procedure produces
solutions of the SSSG equations-of-motion~\eqref{www1} with vanishing
gauge fields.
To start with, using \eqref{hqq} along with \eqref{drs} and
\eqref{oss} one quickly deduces
\EQ{
\partial_\pm\chi(\lambda)\chi(\lambda)^{-1}=\frac{\chi(\mp1)\Lambda_\pm
\chi(\mp1)^{-1}-\chi(\lambda)\Lambda_\pm\chi(\lambda)^{-1}}{1\pm\lambda}
\label{cbb5}
}
from which it follows that
\EQ{
\partial_\pm\chi(\pm1)\chi(\pm1)^{-1}=\tfrac12\big(\chi(\mp1)\Lambda_\pm
\chi(\mp1)^{-1}-\chi(\pm1)\Lambda_\pm\chi(\pm1)^{-1}\big)\ .
\label{cbb1}
}
By writing $\gamma=\newf_0^{-1/2}\chi(+1)^{-1}\chi(-1)\newf_0^{1/2}$, and using
$\partial_\pm \newf_0=\Lambda_\pm \newf_0$, we have
\SP{
\gamma^{-1}\partial_+\gamma=&-\tfrac12\newf_0^{-1/2}\chi(-1)^{-1}\chi(+1)\Lambda_+\chi(+1)^{-1}
\chi(-1)\newf_0^{1/2}+\tfrac12\Lambda_+\\
&-\newf_0^{-1/2}\chi(-1)^{-1}\partial_+\chi(+1)\chi(+1)^{-1}\chi(-1)\newf_0^{1/2}\\
&+\newf_0^{-1/2}\chi(-1)^{-1}\partial_+\chi(-1)\newf_0^{1/2}\ .
}
Using the upper-sign identity \eqref{cbb1}, one sees that the
third term cancels the first two, to leave
\EQ{
\gamma^{-1}\partial_+\gamma=\newf_0^{-1/2}\chi(-1)^{-1}\partial_+\chi(-1)\newf_0^{1/2}\ .
\label{cbb4}
}
Then 
\SP{
\partial_-\big(\gamma^{-1}\partial_+\gamma\big)=&
-\tfrac12\Lambda_-\newf_0^{-1/2}\chi(-1)^{-1}\partial_+\chi(-1)\newf_0^{1/2}\\ &+
\tfrac12\newf_0^{-1/2}\chi(-1)^{-1}\partial_+\chi(-1)\newf_0^{1/2}\Lambda_-\\
&+\newf_0^{-1/2}\chi(-1)^{-1}\partial_+\big(\partial_-\chi(-1)\chi(-1)^{-1}\big)\chi(-1)
\newf_0^{1/2}
\label{cbb2}
}
Next, we use the lower-sign identity \eqref{cbb1} to re-write the third
term as
\SP{
&\partial_+\big(\partial_-\chi(-1)\chi(-1)^{-1}\big)\\ =&
-\tfrac12\partial_+\chi(-1)\Lambda_-\chi(-1)^{-1}+
\tfrac12\chi(-1)\Lambda_-\chi(-1)^{-1}\partial_+\chi(-1)\chi(-1)^{-1}\\
&+\tfrac12\partial_+\chi(+1)\Lambda_-\chi(+1)^{-1}-
\chi(+1)\Lambda_-\chi(+1)^{-1}\partial_+\chi(+1)\chi(+1)^{-1}\ .
}
The first two terms cancel the first two terms in \eqref{cbb2} to leave
\SP{
\partial_-\big(\gamma^{-1}\partial_+\gamma\big)&=\tfrac12\newf_0^{-1/2}\chi(-1)^{-1}
\partial_+\chi(+1)\Lambda_-\chi(+1)^{-1}\chi(-1)\newf_0^{1/2}\\
&-\tfrac12\newf_0^{-1/2}\chi(-1)\chi(+1)\Lambda_-\chi(+1)^{-1}\partial_+\chi(+1)
\chi(+1)^{-1}\chi(-1)\newf_0^{1/2}\ .
}
Finally, we use the upper-sign identity in \eqref{cbb1} again and
the fact that $[\Lambda_+,\Lambda_-]=0$, to end up with
\EQ{
\partial_-\big(\gamma^{-1}\partial_+\gamma\big)=\frac{1}
4[\Lambda_+,\gamma^{-1}\Lambda_-\gamma]\ .
}
This 
is \eqref{www1} with $A_+^{(L)}=A_-^{(R)}=0$.

The next thing to prove is that the constraints \eqref{Constraints} are
satisfied. Taking the
residue of the upper sign in \eqref{cbb5} at $\lambda=-1$, gives
\EQ{
\partial_+\chi(-1)\chi(-1)^{-1}=-\partial_\lambda\chi(-1)\Lambda_+\chi(-1)^{-1}
+\chi(-1)\Lambda_+\chi(-1)^{-1}\partial_\lambda\chi(-1)\chi(-1)^{-1}\
.
}
Substituting this in \eqref{cbb4}, gives
\EQ{
\gamma^{-1}\partial_+
\gamma=-[\newf_0^{-1/2}\chi(-1)^{-1}\partial_\lambda\chi(-1)\newf_0^{1/2},
\Lambda_+\big]\ .
}
with a similar expression for $\partial_-\gamma\gamma^{-1}$. Hence,
$\gamma^{-1}\partial_+\gamma$ and $\partial_-\gamma\gamma^{-1}$ are in the
image of the adjoint action of $\Lambda_\pm$. Then, provided that the orthogonal decompositions~\eqref{Orthogonal} hold, which is always true if the symmetric space has definite signature, the constraints \eqref{Constraints} are satisfied.

\end{document}